\RequirePackage{fix-cm}
\documentclass[twocolumn, epjc3]{svjour3}
\usepackage[square, numbers, comma, sort&compress]{natbib}

\journalname{Eur. Phys. J. A}

\usepackage{latexsym}
\usepackage{amsmath}
\usepackage{amssymb}
\usepackage{amsfonts}

\usepackage[mathscr,scaled=1.15]{urwchancal}
\DeclareFontFamily{OT1}{pzc}{}
\DeclareFontShape{OT1}{pzc}{m}{it}%
{<-> s * [1.15] pzcmi7t}{}
\DeclareMathAlphabet{\mathpzc}{OT1}{pzc}{m}{it}

\usepackage{color}

\usepackage{supertabular}
\usepackage{placeins}
\usepackage{epsfig}
\usepackage{graphicx}

\definecolor{purple}{rgb}{0.5,0,0.5}
\definecolor{blue}{rgb}{0.0,0,0.9}
\definecolor{prdblue}{rgb}{0.133,0.118,0.498}
\usepackage[colorlinks=true, pdfstartview=FitV, linkcolor=prdblue, citecolor= prdblue, urlcolor=prdblue]{hyperref}

\def\tstrut{\vrule height2.25ex depth0pt width0pt} 

\begin{document}

\title{$\,$\\[-7ex]\hspace*{\fill}{\normalsize{\sf\emph{Preprint no}. NJU-INP 046/21}}\\[1ex]
Dynamical diquarks in the ${\boldsymbol{\gamma^{(\ast)} p\to N(1535)\tfrac{1}{2}^-}}$ transition
}

\author{K.~Raya\thanksref{NKU,UNAM}
       \and
       L.\,X.~Guti\'errez-Guerrero\thanksref{MCTP}
        \and
       A.~Bashir\thanksref{UMich} 
       \and
       L.~Chang\thanksref{NKU} 
       \and
       Z.-F.~Cui\thanksref{NJU,INP}
       \and
       Y.~Lu\thanksref{NJU,INP}
       \and
       C.\,D. Roberts\thanksref{NJU,INP}
       \and
       J.~Segovia\thanksref{UPO,INP}
}

\authorrunning{K.\ Raya \emph{et al}.} 

\institute{School of Physics, Nankai University, Tianjin 300071, China \label{NKU}
\and
Instituto de Ciencias Nucleares, Universidad Nacional Aut\'onoma de M\'exico, Apartado Postal 70-543, CDMX 04510, M\'exico \label{UNAM}
            \and
            CONACyT-Mesoamerican Centre for Theoretical Physics, Universidad Aut\'onoma de Chiapas, \\ \mbox{$\;\;$}Carretera Zapata Km.\ 4, Real del Bosque (Ter\'an), Tuxtla Guti\'errez 29040, Chiapas, M\'exico \label{MCTP}
            \and
            Instituto de F\'{i}sica y Matem\'aticas, Universidad Michoacana de San Nicol\'as de Hidalgo, \\ \mbox{$\;\;$}Edificio C-3, Ciudad Universitaria, C.P. 58040,  Morelia,  Michoac\'an, M\'exico \label{UMich}
            \and
            School of Physics, Nanjing University, Nanjing, Jiangsu 210093, China \label{NJU}
           \and
           Institute for Nonperturbative Physics, Nanjing University, Nanjing, Jiangsu 210093, China \label{INP}
           \and
           Dpto. Sistemas F\'isicos, Qu\'imicos y Naturales, Univ.\ Pablo de Olavide, E-41013 Sevilla, Spain \label{UPO}\\[1ex]
Email:
\href{mailto:khepani@nankai.edu.cn}{khepani@nankai.edu.cn} (K.\ Raya);
\href{mailto:lxgutierrez@mctp.mx}{lxgutierrez@mctp.mx} (L.\,X.\ Guti\'errez-Guerrero);\\
\hspace*{1em}\href{mailto:adnan.bashir@umich.mx}{adnan.bashir@umich.mx} (A.\ Bashir);
\href{mailto:leichang@nankai.edu.cn}{leichang@nankai.edu.cn} (L.\ Chang);
\href{mailto:phycui@nju.edu.cn}{phycui@nju.edu.cn} (Z.-F.\ Cui);\\
\hspace*{1em}\href{mailto:luya@nju.edu.cn}{luya@nju.edu.cn} (Y. Lu);
\href{mailto:cdroberts@nju.edu.cn}{cdroberts@nju.edu.cn} (C.\,D.\ Roberts);
\href{mailto:jsegovia@upo.es}{jsegovia@upo.es} (J.\ Segovia)
            }

\date{2021 August 04}

\maketitle



\begin{abstract}
The $\gamma^{(\ast)}+p \to N(1535) \tfrac{1}{2}^-$ transition is studied using a symmetry-preserving regularisation of a vector$\,\otimes\,$vector contact interaction (SCI).  The framework employs a Poincar\'e-covariant Faddeev equation to describe the initial and final state baryons as quark+di\-quark composites, wherein the diquark correlations are fully dynamical, interacting with the photon as allowed by their quantum numbers and continually engaging in breakup and recombination as required by the Faddeev kernel.  The presence of such correlations owes largely to the mechanisms responsible for the emergence of hadron mass; and whereas the nucleon Faddeev amplitude is dominated by scalar and axial-vector diquark correlations, the amplitude of its parity partner, the $N(1535) \tfrac{1}{2}^-$, also contains sizeable pseudoscalar and vector diquark components.  It is found that the $\gamma^{(\ast)}+p \to N(1535) \tfrac{1}{2}^-$ helicity amplitudes and related Dirac and Pauli form factors are keenly sensitive to the relative strengths of these diquark components in the baryon amplitudes, indicating that such resonance electrocouplings possess great sensitivity to baryon structural details.  Whilst SCI analyses have their limitations, they also have the virtue of algebraic simplicity and a proven ability to reveal insights that can be used to inform more sophisticated studies in frameworks with closer ties to quantum chromodynamics.
\end{abstract}

\maketitle


\section{Introduction}
\label{Introduction}
Experiments at modern facilities have provided a great deal of information about nucleon structure and more is anticipated \cite{Aznauryan:2012ba, Brodsky:2020vco, Carman:2020qmb, Barabanov:2020jvn, Chen:2020ijn, Arrington:2021biu}.  Such data is crucial because it is notoriously difficult for theory to deliver \emph{ab initio} hadron structure predictions.  First, the relevant bound-state problems in quantum chromodynamics (QCD) must be solved; then the associated scattering problems involving the composite bound-state solutions.  Only precise data can decide whether the solutions are sound.

To further confound progress, the nucleon is only the ground state of the QCD Hamiltonian.  In attempting to develop insights into hadron structure, model Hamiltonians have typically been used.  The issue here is that the ground state is just one isolated member of a set of Hamiltonian eigenvectors with a countable infinity of elements: many Hamiltonians can possess practically identical ground states and yet produce excited-state spectra that are vastly different.  Moreover, masses alone, being infrared-dominated quantities, contain relatively little information.  Distinct Hamiltonians can satisfactorily reproduce known hadron spectra; but those same Hamiltonians may deliver predictions that disagree markedly when employed to compute structural properties \cite{Brodsky:2020vco, Carman:2020qmb, Barabanov:2020jvn}.  Such properties -- like the $Q^2$-depen\-dence of elastic and transition form factors -- possess the greatest discriminating power.   Hence, a sure way to develop deeper understanding is for theory to compute these observables.

A clear case is provided by the Roper resonance, $N(1440)\tfrac{1}{2}^+$.  From discovery \cite{Roper:1964zza, BAREYRE1964137, AUVIL196476, PhysRevLett.13.555, PhysRev.138.B190} until the turn of the current millennium, its nature was the source of great puzzlement.  This changed following the collection and analysis of precise electroproduction data to $W=2\,$GeV and $Q^2=4.5\,$GeV$^2$ \cite{Aznauryan:2008pe, Aznauryan:2009mx, Aznauryan:2011qj, Mokeev:2012vsa, Mokeev:2015lda}.  Today it is recognised that the Roper is, at heart, the first radial excitation of the nucleon \cite{Burkert:2019bhp, Sun:2019aem}.  A wide-ranging effort, involving many research arms, revealed that the Roper consists of a well-defined dressed-quark core, which influences the system's properties at all length-scales, but is dominant when the resonance is probed with $Q^2\gtrsim m_N^2$, where $m_N$ is the nucleon mass.  That core is augmented by a meson cloud, which both reduces the Roper's core mass by $\approx 20$\% and, at low-$Q^2$, contributes an amount to the electroproduction transition form factors that is comparable in magnitude with that of the dressed-quark core, but vanishes rapidly as $Q^2$ is increased beyond $m_N^2$.

The next simplest excited state of the nucleon is the $N(1535)\tfrac{1}{2}^-$ and it is natural to ask how these two systems are related.  In constituent-quark models, the $N(1535)\tfrac{1}{2}^-$ is pictured as a $P$-wave excitation of the nucleon \cite{Isgur:1978xj}, \emph{i.e}.\ a member of the $(70,1_1^-)$ supermultiplet of $SU(3)\otimes O(3)$, with $L=1$ and constituent-quark total spin $S=1/2$, coupled to $J=L+S=1/2$.  However, QCD is a relativistic quantum field theory, in which case $L$ and $S$ are not good quantum numbers.  Even if they were, owing to the loss of particle number conservation, it is not clear \emph{a priori} just with which degrees-of-freedom $L$, $S$ should be connected.  This issue is related to the fact that the constituent-quarks used in building quantum mechanical models have no known mathematical connection with QCD.

This importance of the $N(1535)\tfrac{1}{2}^-$ is heightened further by the fact that, in a symmetry-preserving treatment using relativistic quantum field theory, one may generate the interpolating field for the parity partner of any given state via a chiral rotation of that associated with the original state.  It follows that parity partners will be degenerate in mass and alike in structure in all theories that possess a chiral symmetry realised in the Wigner-Weyl mode.
Such knowledge has long made the mass-splittings between strong-interaction parity partners a subject of interest.  A well known example is provided by the $\rho(770)$- and $a_1(1260)$-mesons: viewed as chiral and hence parity partners, it has been argued \cite{Weinberg:1967kj} that their mass and structural differences owe entirely to dynamical chiral symmetry breaking (DCSB), \emph{viz}.\ realisation of chiral symmetry in the Nambu-Goldstone mode.

DCSB is a material corollary of the emergence of hadron mass (EHM); hence, linked closely with confinement \cite{Roberts:2020hiw, Roberts:2020udq}.  Regarding DCSB's role in explaining the splittings between parity partners, additional insights have been provided by studies of the bound-state equations appropriate to the $\rho$- and $a_1$-mesons.  In their rest frames, one finds their Poincar\'e-covariant wave functions are chiefly $S$-wave in nature \cite{Chang:2011ei, Eichmann:2016yit, Qin:2020jig}, even though both possess nonzero angular momentum \cite{Gao:2014bca}, whose magnitude influences the size of the splitting \cite{Chang:2011ei, Eichmann:2016yit, Qin:2020jig}.

$N(1535)\tfrac{1}{2}^-$\,structure has similarly been studied, with the Poincar\'e-covariant Faddeev equation introduced in Refs.\,\cite{Cahill:1988dx, Burden:1988dt, Reinhardt:1989rw, Efimov:1990uz} being employed to compute the mass and Faddeev amplitude of this system for comparison with that of the nucleon \cite{Chen:2017pse}.  The efficacy of this Faddeev equation approach is grounded on the existence of nonpointlike, electromagnetically-active quark+quark (diquark) correlations within all baryons \cite{Barabanov:2020jvn}, whose appearance is a consequence of EHM.  Such correlations exist in all channels: scalar, pseudovector, pseudoscalar and vector, with effective masses growing in the order listed \cite{Lu:2017cln, Chen:2017pse, Yin:2021uom, Eichmann:2016hgl}.  In the $J^P=\tfrac{1}{2}^+$ nucleon and Roper, scalar and pseudovector diquarks are overwhelmingly dominant; and the associated rest-frame wave functions are largely $S$-wave in nature.
On the other hand, the $N(1535)\tfrac{1}{2}^-$ fits a different picture \cite{Lu:2017cln, Chen:2017pse, Yin:2021uom}: a fair estimate of its mass is obtained by retaining only pseudovector diquarks; the amplitudes describing the dressed-quark core contain roughly equal fractions of even- and odd-parity diquarks; and the associated rest-frame wave functions are predominantly $P$-wave in nature, but possess measurable $S$-wave components.  These structural predictions can be tested by comparing the entailed $\gamma^{(\ast)} p\to N(1535)\tfrac{1}{2}^-$ transition form factors with existing data \cite{Dugger:2009pn, Aznauryan:2009mx}.

Here it is worth relating some pertinent features of available data on the $\gamma^{(\ast)} N\to N(1535)\tfrac{1}{2}^-$ transition.  Pion electroproduction results are available to $Q^2=4.5\,$GeV$^2$ \cite{Dugger:2009pn, Aznauryan:2009mx}, with sufficient precision to enable extraction of both the transverse ($A_{1/2}$) and longitudinal ($S_{1/2}$) helicity amplitudes.  The $Q^2$ dependence of $A_{1/2}$ confirms that found earlier in $\eta$ electroproduction \cite{Armstrong:1998wg, Thompson:2000by, Aznauryan:2003zg, Denizli:2007tq}.  Importantly, whilst $S_{1/2}$ could not be obtained from the $\eta$ data, owing to lack of precision, the $S_{1/2}$ results from $\pi$ electroproduction have provided a real test for theory, with quark models typically failing to reproduce the sign \cite{Eichmann:2018ytt}.  A similar failing of quark models was also found with the Roper resonance \cite{Burkert:2019bhp}.
Additional information relating to quark model studies may be found elsewhere \cite{Capstick:2000qj, Crede:2013kia, Giannini:2015zia}.

Herein, motivated by the above considerations, we present a calculation of the $\gamma^{(\ast)} p\to N(1535)\tfrac{1}{2}^-$ transition form factors using a confining, symmetry-preserving regularisation of a vector$\,\otimes\,$vector contact interaction (SCI) \cite{GutierrezGuerrero:2010md}.  This framework has the merit of providing a largely algebraic solution to the problem, which makes it ideal for developing insights that will be useful to the more sophisticated studies that must follow.  Such was the case for Roper resonance electroproduction \cite{Wilson:2011aa, Segovia:2015hra}.

Our contribution is prepared as follows.
Section~\ref{sec:Theory}, augmented by appendices, introduces the SCI and its application to baryon elastic and transition form factors, including descriptions of the Faddeev equation and electromagnetic interaction current.
Low-$Q^2$ properties of the nucleon and $N(1535)$ elastic form factors are reported in Sec.\,\ref{ElasticFF}; and the $\gamma^{(\ast)} p\to N(1535)\tfrac{1}{2}^-$ helicity amplitudes and transition form factors are discussed in Secs.\,\ref{sec:AAs}\,--\,\ref{sec:Dissections}.
Section~\ref{Epilogue} is a summary and outlook.

\section{Theoretical framework}\label{sec:Theory}
\unskip
\subsection{Quark-quark interaction}
\label{SecSCI}
Our starting point is a statement of the quark-quark interaction.  In QCD, this is now known with some certainty \cite{Binosi:2014aea, Deur:2016tte, Cui:2019dwv}, as are its consequences: whilst the effective charge, and gluon and quark masses run with momentum-squared, $k^2$, they all saturate at infrared momenta, each changing by $\lesssim20$\% on $0\lesssim \surd k^2 \lesssim m_0 \approx m_p/2$, where $m_0$ is a renormalisation-group-invariant gluon mass-scale and $m_p$ is the proton mass.  It follows that, employed judiciously, the SCI can provide insights and useful results for a diverse array of observables \cite{Farias:2006cs, GutierrezGuerrero:2010md, Roberts:2010rn, Roberts:2011cf, Roberts:2011wy, Wilson:2011aa, Chen:2012qr, Pitschmann:2012by, Segovia:2013uga, Xu:2015kta, Bedolla:2015mpa, Bedolla:2016yxq, Serna:2016ifh, Lu:2017cln, Yin:2021uom, Xu:2021iwv}.

\begin{table}[t]
\caption{\label{tabledressedquark}
Computed dressed-quark properties, required as input for the bound-state equations employed herein.  All results obtained with contact-interaction strength $\alpha_{\rm IR} =0.93 \pi$, and (in GeV) infrared and ultraviolet regularisation scales $\Lambda_{\rm ir} = 0.24=1/r_{\rm ir}$, $\Lambda_{\rm uv}=0.905=1/r_{\rm uv}$.  N.B.\ These parameters take the values determined in the spectrum calculation of Ref.\,\protect\cite{Chen:2012qr}, we assume isospin symmetry throughout, and $\Lambda_{\rm ir}>0$ implements dressed-quark confinement \cite{Ebert:1996vx}.
(All dimensioned quantities are listed in GeV.  Related values of $s$-quark masses are listed so as to provide additional context.)}
\begin{center}
\begin{tabular*}
{\hsize}
{
c@{\extracolsep{0ptplus1fil}}
c@{\extracolsep{0ptplus1fil}}
c@{\extracolsep{0ptplus1fil}}
c@{\extracolsep{0ptplus1fil}}
c@{\extracolsep{0ptplus1fil}}
c@{\extracolsep{0ptplus1fil}}
c@{\extracolsep{0ptplus1fil}}
c@{\extracolsep{0ptplus1fil}}}\hline
\multicolumn{4}{c}{input: current masses} & \multicolumn{4}{c}{output: dressed masses}\\
$m_0$ & $m_u$ & $m_s$ & $m_s/m_u$ & $M_0$ &   $M_u$ & $M_s$ & $M_s/M_u$  \\\hline
0 & 0.007  & 0.17 & 24.3 & 0.36 & 0.37 & 0.53 & 1.43  \\\hline
\end{tabular*}
\end{center}
\end{table}

Our SCI approach to baryons is detailed in Refs.\,\cite{Roberts:2011cf, Lu:2017cln}.  It is based upon the rainbow-ladder (RL) approximation to those equations in quantum field theory that are directly involved in formulating the three-body problem \cite{Binosi:2016rxz}.  In addition to the light-quark current mass, the SCI is specified by three parameters: interaction strength, $\alpha_{\rm IR}$, and infrared and ultraviolet cutoffs $\Lambda_{\rm ir}$, $\Lambda_{\rm uv}$.  They are listed in Table~\ref{tabledressedquark} along with the results they yield for the masses of the dressed $u=d$- and $s$-quarks when used in the gap equation.

\subsection{Faddeev equation}\label{subsec:BoundState}
\begin{figure}[t]
\centerline{%
\includegraphics[clip, width=0.45\textwidth]{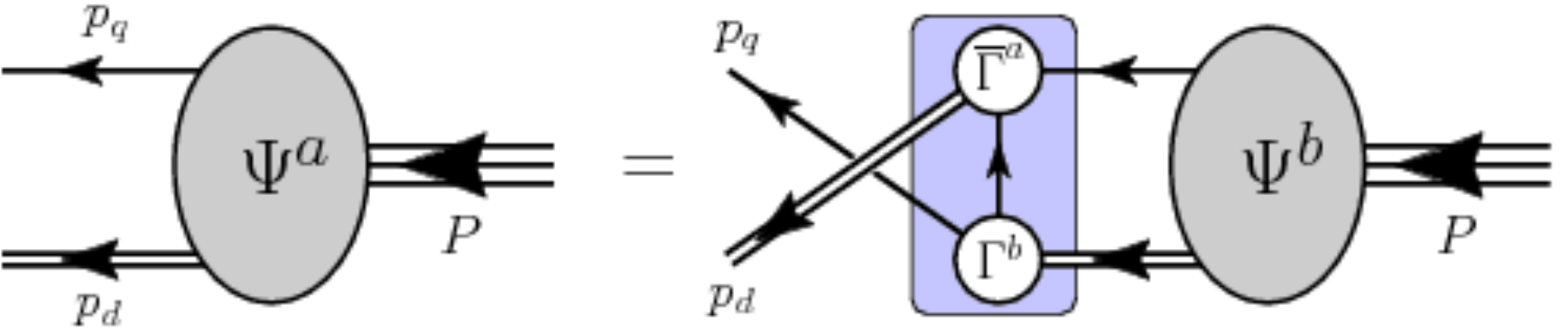}}
\caption{\label{figFaddeev}
Baryon = quark+diquark Faddeev equation: a linear integral equation for the Poincar\'e-covariant matrix-valued function $\Psi$, the Faddeev amplitude for a state with total momentum $P= p_q + p_d$.  It describes the relative momentum correlation between the dressed-quarks and -diquarks.  Legend. Shaded rectangle -- kernel of the Faddeev equation; \emph{single line} -- dressed-quark propagator; $\Gamma$ -- diquark correlation amplitude; and \emph{double line} -- diquark propagator.
For light-quark baryons, active diquark correlations exist in the following channels: isoscalar-scalar -- $[ud]_{0^+}$; isovector-pseudovector -- $\{dd\}_{1^+}$, $\{ud\}_{1^+}$, $\{uu\}_{1^+}$; isoscalar-pseudoscalar -- $[ud]_{0^-}$; and isoscalar-vector -- $[ud]_{1-}$.
%
%
}
\end{figure}

Following Refs.\,\cite{Cahill:1988dx, Burden:1988dt, Cahill:1988zi, Reinhardt:1989rw, Efimov:1990uz}, we consider baryons to be described by the Faddeev equation depicted in Fig.\,\ref{figFaddeev}.   The derivation of this equation is grounded on an important corollary of EHM; namely, any interaction capable of creating pseudo--Nambu-Goldstone modes as bound-states of a light dressed-quark and -antiquark, and reproducing the measured value of their leptonic decay constants, will necessarily, \emph{inter alia}, also generate strong colour-antitriplet correlations between any two dressed-quarks contained within a baryon \cite{Barabanov:2020jvn}.   

The properties of such diquarks are known.  As co\-lour-carrying correlations, diquarks are confined \cite{Bhagwat:2004hn}.  Moreover, a diquark with spin-parity $J^P$ may be viewed as a partner to the analogous $J^{-P}$ meson \cite{Cahill:1987qr}.  Hence, focusing on light-quark systems, the strongest diquark correlations are isoscalar-scalar, $[ud]_{0^+}$; and iso\-vector-pseudo\-vector, $\{dd\}_{1^+}$, $\{ud\}_{1^+}$, $\{uu\}_{1^+}$.  Isoscalar-pseu\-doscalar, $[ud]_{0^-}$, and isoscalar-vector, $[ud]_{1-}$, diquark correlations also exist and play a role in negative-parity baryons \cite{Eichmann:2016yit, Eichmann:2016hgl, Lu:2017cln, Chen:2017pse, Yin:2021uom}.  
The SCI does not support an isovec\-tor-vector correlation.  Furthermore, it is typically found to be a very weak correlation using any interaction \cite{Eichmann:2016yit, Eichmann:2016hgl, Lu:2017cln, Chen:2017pse}; hence, plays no material role in any system studied thus far.
Whilst no pole-masses exist, the following mass-scales, which express the strength and range of the correlation, may be associated with the diquarks (in GeV):
\begin{equation}
\label{diquarkmasses}
\begin{array}{c|c|c|c}
m_{[ud]_{0^+}} & m_{\{uu\}_{1^+}}  & m_{[ud]_{0^-}} & m_{[ud]_{1-}} \\\hline
0.78 & 1.06 & 1.15 & 1.33
\end{array}\,.
\end{equation}
The values in Eq.\,\eqref{diquarkmasses} are SCI predictions \cite{Lu:2017cln}, obtained using the parameters described in Table~\ref{tabledressedquark}.  In the isospin symmetry limit, $m_{\{dd\}_{1^+}} = m_{\{ud\}_{1^+}} = m_{\{uu\}_{1^+}}$.

Given that the diquark concept continues to be view\-ed differently by various practitioners, we emphasise that the diquark correlations which play a key role in our study are vastly different from the static, pointlike ``diquarks'' introduced originally \cite{Anselmino:1992vg} in an attempt to solve the so-called ``missing resonance'' problem \cite{Aznauryan:2011ub}, \emph{i.e}.\ the fact that quark models predict many more baryon states than were observed in the previous millennium \cite{Burkert:2004sk}.  The diquarks in Fig.\,\ref{figFaddeev} are fully dynamical: they appear in the Faddeev kernel, which requires their continual breakup and reformation; and matching indications from lQCD \cite{Edwards:2011jj}, baryon spectra generated by this Faddeev equation are far richer than those obtained using any two-body model.  Additionally, e.g.\ the presence of such diquark correlations within baryons enforces distinct interaction patterns for the singly- and doubly-represented valence-quarks within the proton and its excited states, as exhibited elsewhere \cite{Roberts:2013mja, Segovia:2015ufa, Segovia:2016zyc, Cui:2020rmu, Chen:2020ijn, Roberts:2020hiw}.

As just noted, the kernel in Fig.\,\ref{figFaddeev} involves diquark breakup and reformation via exchange of a dressed-quark.  In order to present a transparent analysis, we follow Refs.\,\cite{Roberts:2011cf, Chen:2012qr} and introduce a simplification, \emph{viz}.\ quark propagation between the diquarks is represented as
\begin{equation}
S(k) \to \frac{g_8^2}{M_u}\,,
\label{staticexchange}
\end{equation}
where $g_8$ is a coupling constant.  This is a variant of the  ``static approximation'' introduced in Ref.\,\cite{Buck:1992wz}.  It has a marked impact on the Faddeev amplitudes, forcing them to be momentum-independent, just like the diquark Bethe-Salpeter amplitudes; but calculations reveal that it has little impact on the computed masses \cite{Xu:2015kta}.  The value
$g_{8}=1.18$,
was fixed in Ref.\,\cite{Roberts:2011cf} in order to produce quark-core masses for the nucleon and $\Delta$-baryon that are each inflated by roughly 0.2\,GeV so that the experimental values are reproduced after meson-baryon final-state interactions are incorporated \cite{Hecht:2002ej, Eichmann:2008ae, Eichmann:2008ef, Suzuki:2009nj, Kamano:2013iva, Segovia:2014aza, Segovia:2015hra}.

One can now construct the Faddeev kernels for the $J^P=\tfrac{1}{2}^+$ proton and $J^P=\tfrac{1}{2}^-$ $N(1535)$.  Their structure depends on the form of the associated Faddeev amplitudes; and assuming the latter is the nucleon's parity partner, then
\begin{align}
\nonumber
\Psi^\pm(P)= \psi^\pm & u(P)  =  \Gamma^1_{0^+} \Delta^{0^+}(K)\, {\mathpzc S}^\pm(P)u(P)   \\
\nonumber
& + \mbox{$\sum$}_{{\mathpzc j}=1,2}\Gamma^{\mathpzc j}_{1^+\mu }\Delta^{1^+}_{\mu\nu}(K) {\mathpzc A}_\nu^{\pm {\mathpzc j}}(P) u(P)\\ %
\nonumber
& + \Gamma^1_{0^-}(K) \Delta^{0^-}(K) {\mathpzc P}^{\pm}(P)\,u(P)  \\
& + \Gamma^{1}_{1^-\mu } \Delta^{1^-}_{\mu\nu}(K) {\mathpzc V}^\pm_\nu(P) u(P)\,,
\label{nucleonamplitude}
\end{align}
where $u(P)$ is a Dirac spinor for the on-shell baryon;
$\Delta^{0^+}(K)$, etc., are standard propagators for scalar or vector bosons, detailed in Refs.\,\cite{Roberts:2011cf, Chen:2012qr}, with the appropriate masses from Eq.\,\eqref{diquarkmasses};
${\mathpzc j}=1,2$ means $\{uu\}_{1^+}$, $\{ud\}_{1^+}$;
and, with $\hat P^2=-1$, $\mathpzc{G}^{+(-)} = \mathbf{I}_{\rm D} (\gamma_5)$,
\begin{align}
\nonumber
{\mathpzc S}^\pm & = \mathpzc{s}^\pm\,\mathbf{I}_{\rm D} \mathpzc{G}^\pm \,, \quad
i {\mathpzc P}^\pm  = \mathpzc{p}^\pm\,  \gamma_5 \mathpzc{G}^\pm\,,
\\
i{\mathpzc A}_\mu^{\pm {\mathpzc j}} & = (\mathpzc{a}_1^{\pm {\mathpzc j}} \gamma_5\gamma_\mu - i \mathpzc{a}_2^{\pm {\mathpzc j}} \gamma_5 \hat P_\mu) \mathpzc{G}^\pm \,, \label{spav} \\
\nonumber
i \mathpzc{V}_\mu^\pm & = (\mathpzc{v}_1^\pm \gamma_\mu - i \mathpzc{v}_2^\pm \mathbf{I}_{\rm D}  \hat P_\mu)\gamma_5\mathpzc{G}^\pm\,.
\end{align}

The masses, $m^2_\pm$, and eigenvectors $(\mathpzc{s}^\pm$, $\mathpzc{a}_1^{\pm {\mathpzc j}}$, $\mathpzc{a}_2^{\pm {\mathpzc j}}$, $\mathpzc{p}^\pm,\mathpzc{v}_1^{\pm}, \mathpzc{v}_2^{\pm})$, can now be obtained by substituting the amplitudes from Eq.\,\eqref{nucleonamplitude} into the Faddeev equation depicted in Fig.\,\ref{figFaddeev} and solving the resulting eigenvalue problems.  Owing to isospin symmetry in the two cases considered, the kernel can be reduced to a $6\times 6$ matrix because $\mathpzc{a}^{\pm 2} = - \mathpzc{a}^{\pm 1}/\surd 2$.  Following the procedure detailed in Ref.\,\cite{Lu:2017cln}, using $g_{\rm DB}=0.2$ as the value for the in-baryon spin-orbit-repulsion parameter, the results are (in GeV):
\begin{subequations}
\begin{align}
\label{BothMasses}
& m_{N(940)} = 1.14\,,\, \quad m_{N(1535)} = 1.73 \,, \\
& \begin{array}{l|ccc|ccc}\hline
\mbox{\rm baryon} & s & a_1^1 & a_2^1 & p & v_1 & v_2 \\
\hline
\phantom{1}N(940)\tfrac{1}{2}^+  & 0.88 & 0.38 & -0.06 & 0.02 & \phantom{-}0.02 & 0.00 \\
N(1535)\tfrac{1}{2}^-
    & 0.66 & 0.20 & \phantom{-}0.14 & 0.68 & \phantom{-}0.11 & 0.09 \\
\end{array}
\label{eq:amplitudes}
\end{align}
\end{subequations}
Evidently, as noted in the Introduction, scalar and pseudovector diquarks dominate in the nucleon whereas the pseudoscalar diquark is prominent in the $N(1535)$, albeit in the presence of a significant scalar diquark component.

If one varies $g_{\rm DB} \to g_{\rm DB} (1 \pm 0.5 )$, then
$m_{N(1535)}$ $= (1.67, 1.82)\,$GeV and
\begin{equation}
\begin{array}{c|ccc|ccc}
N(1535)\tfrac{1}{2}^- & s & a_1^1 & a_2^1 & p & v_1 & v_2 \\
\hline
g_{\rm DB} \, 1.5 & 0.76 & 0.27 & 0.18 & 0.49 & \phantom{-}0.12 & 0.08 \\ 
g_{\rm DB} \, 1.0 &  0.66 & 0.20 & 0.14 & 0.68 & \phantom{-}0.11 & 0.09 \\
g_{\rm DB}\, 0.5    & 0.35 & 0.04 & 0.00 & 0.92 & -0.05 & 0.18 \\
\end{array}\,,
\label{eq:amplitudesgDB}
\end{equation}
where we have here repeated the $N(1535)$ result from Eq.\,\eqref{eq:amplitudes} as the middle row so as to simplify comparisons.
More realistic interactions deliver qualitatively similar weightings and similar sensitivity to the strength of $g_{\rm DB}$ \cite{Chen:2017pse}.

%
The empirical masses of the nucleon and its parity partner are (GeV) \cite{Zyla:2020zbs}: $0.939$ and $1.51 - i \, 0.07$.  At first glance, these values seem unrelated to those in Eq.\,\eqref{BothMasses}.  Recall, therefore, that the kernel in Fig.\,\ref{figFaddeev} omits all resonant contributions which may be associated with the meson-baryon final-state interactions that are resummed in dynamical coupled-channels (DCC) models \cite{Suzuki:2009nj, Kamano:2013iva} so as to transform a bare-baryon into the observed state.  Hence, our Faddeev equation should be understood as producing the dressed-quark core of the bound-state, not the completely-dressed object.  In this case it is notable that the results in Eq.\,\eqref{BothMasses} compare favourably with the bare masses determined in DCC models \cite{Suzuki:2009nj}.

\subsection{Photon-baryon interactions}
Three matrix-valued electromagnetic vertices must be considered herein.
The first two are associated with the nucleon and $N(1535)$ elastic form factors, which take the form
\begin{align}
\nonumber
\Gamma_\mu^\pm(P_f,P_i) & =  i e\,
\Lambda_+^\pm (P_f)
\left[ \gamma_\mu F_{1}^{\pm}(Q^2) \right. \\
& \qquad \left. + \frac{1}{2 m_{\pm}} \sigma_{\mu\nu} Q_\nu F_{2}^{\pm}(Q^2)\right] \Lambda_+^\pm(P_i)\,,
\label{Belastic}
\end{align}
where
$e$ is the positron charge;
$\Lambda_+^\pm(P) = \mathpzc{G}^\pm \Lambda_+ (P) \mathpzc{G}^\pm$, with
$\Lambda_+(P)=(m-i\gamma\cdot P)/(2m)$ for a baryon with mass $m$;
$(\pm)=N(940)$, $N(1535)$, respectively; and $Q=P_f - P_i$.
The third vertex is that expressing the $\gamma^{(\ast)} p\to N(1535)\tfrac{1}{2}^-$ transition form factors [$\gamma_\mu^T = \gamma_\mu - \gamma\cdot Q Q_\mu/Q^2$],
\begin{align}
\nonumber
\Gamma_\mu^\ast(P_f,P_i) & =  i e\,  \Lambda_+^- (P_f) \left[ \gamma_\mu^T F_{1}^{\ast}(Q^2) \right. \\
& \left. + \frac{1}{m_++m_{-}} \sigma_{\mu\nu} Q_\nu F_{2}^{\ast}(Q^2)\right] \Lambda_+^+(P_i).\rule{1em}{0ex}
\label{N1535transition}
\end{align}
\emph{N.B}.\ 
Eq.\,(\ref{Belastic}) may be viewed as a special case of Eq.\,(\ref{N1535transition}), simplified by the on-shell condition $\bar u(P_f) \gamma\cdot Q u(P_i) = 0$, valid for elastic processes.

The kinematic constraints are plain.  For the elastic currents,
\begin{align}
P_f^2 = -m_{\pm}^2 = P_i^2 \,,\quad Q^2+ 2 P_i\cdot Q = 0\,;
\end{align}
whereas for the transition current, writing $2 K=P_f+P_i$,
\begin{subequations}
\begin{align}
P_f\cdot P_i & = K^2 - \tfrac{1}{4}Q^2 \\
K\cdot Q &= \frac{m_{+}^2-m_-^2}{2} \,, \\
K^2 &= -\frac{m_{+}^2+m_-^2}{2}-\tfrac{1}{4} Q^2.
\end{align}
\end{subequations}
Our Euclidean metric conventions are detailed in Ref.\,\cite[Appendix\,B]{Segovia:2014aza}.

\begin{figure}[t]
\includegraphics[clip,width=0.33\textwidth]{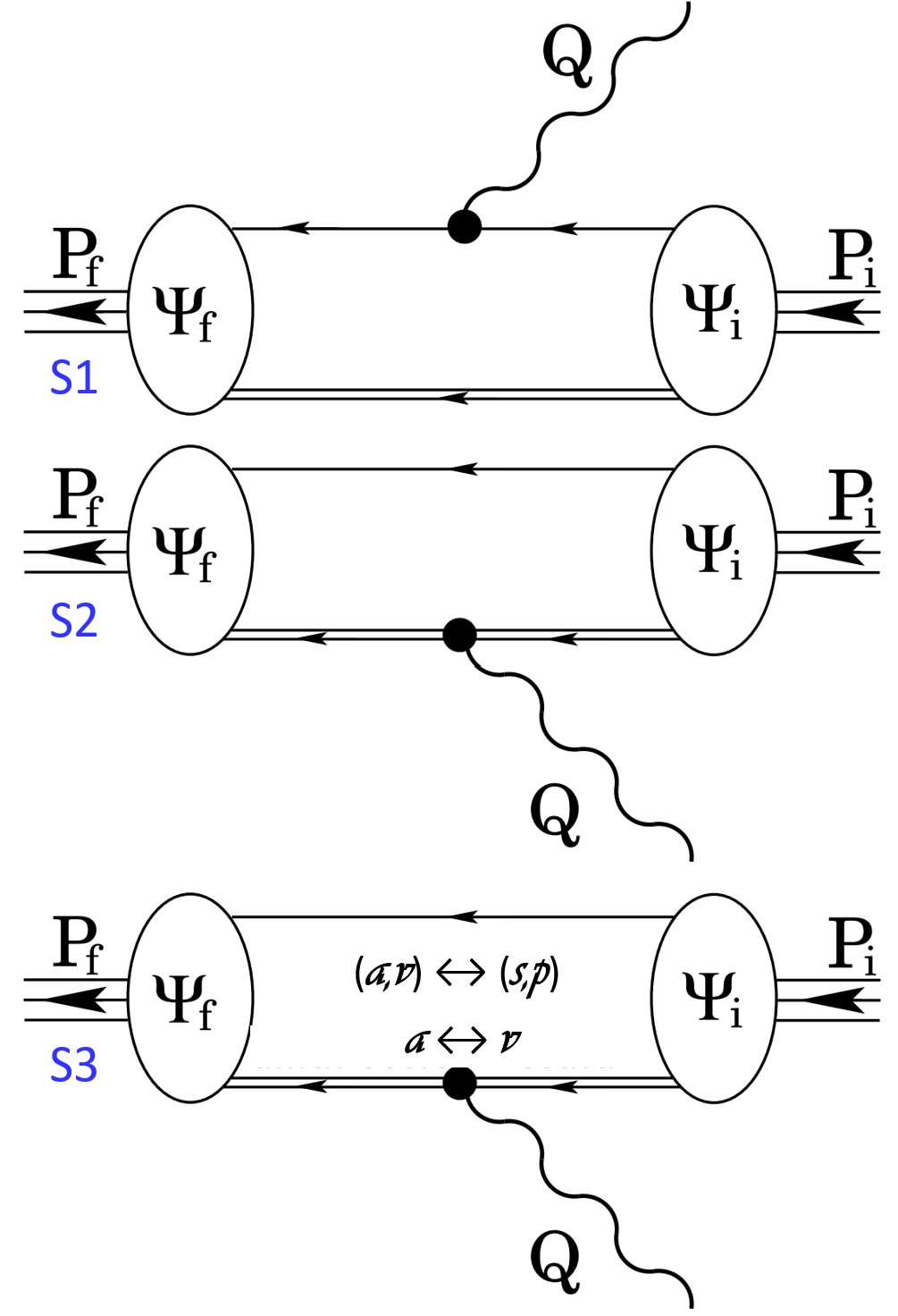}
\caption{\label{fig:current} Interaction vertex which ensures a conserved current for the elastic and transition form factors in Eqs.\,(\protect\ref{Belastic}), (\protect\ref{N1535transition}).  The single line represents the dressed-quark propagator, $S(p)$; the double line, the diquark propagators; and the vertices are described in \protect\ref{App:current}.  From top to bottom: S1 -- photon couples directly to the dressed-quark; S2 -- photon couples to a diquark, in an elastic scattering event; and S3 -- photon induces a transition between different diquarks (axial-vector,vector)$\,\leftrightarrow\,$(scalar,pseudoscalar) and axial-vector$\,\leftrightarrow\,$vector.}
\end{figure}

Using the SCI along with Eq.\,\eqref{staticexchange}, there are three distinct types of contribution to the vertices, Eqs.\,\eqref{Belastic}, \eqref{N1535transition}, \emph{viz}.\
\begin{align}
\nonumber
& \Gamma_\mu^{BA}(P_f,P_i) \\
& = \sum_{I=S1,S2,S3}\int \frac{d^4 l}{(2\pi)^4} \Lambda_+^B(P_f) \Lambda_\mu^I(l;P_f,P_i) \Lambda_+^A(P_i)\,,
\label{JNNastExplicit}
\end{align}
where $BA= ++$, $--$, $-+$.  The individual contributions are illustrated in Fig.\,\ref{fig:current} and detailed in \ref{App:current}.  (Eschewing Eq.\,\eqref{staticexchange}, there is an additional, two-loop contribution \cite{Xu:2015kta}.)

The elastic and transition form factors, Dirac and Pauli, can now be obtained via suitably chosen Dirac-matrix projections of $\Gamma_\mu^{BA}(P_f,P_i)$.  Each yields a weigh\-ted sum of scalar integral contributions from the terms on the right-hand-side of Eq.\,\eqref{JNNastExplicit}, all of which can be evaluated using standard techniques.  The same two projection operators work in every case.

\section{Elastic Form Factors}
\label{ElasticFF}
When planning the calculation of transition form factors, one must first compute the low-$Q^2$ behaviour of the elastic form factors for each of the states involved because: the associated values of $F_1^{\pm}(Q^2=0)$ for the charged states in an isospin multiplet fix the canonical normalisation of the transition; and computing\linebreak $F_2^{\pm}(Q^2\simeq 0)$ costs little additional effort.
The SCI delivers the results in Table~\ref{tabstatic}, with the radii defined via:
\begin{equation}
r_{F}^2 = -\frac{6}{\mathpzc n}\left.\frac{d}{dQ^2} F(Q^2)\right|_{Q^2=0},
\end{equation}
where $F=F_1^\pm,F_2^\pm$; ${\mathpzc n} = F(0)$ if this quantity is nonzero, otherwise ${\mathpzc n} = -1$.
Since all parameters in the SCI were fixed elsewhere \cite{Roberts:2011cf, Chen:2012qr}, these values are predictions.
The uncertainty estimate reflects a variation of the dressed-quark anomalous magnetic moment\linebreak (DqAMM), Eq.\,\eqref{DQAMM}, within the range $0\leq \zeta\leq 0.5$ around the optimal value $\zeta=1/3$ determined as described elsewhere \cite{Wilson:2011aa}.

\begin{table}[!t]
\caption{\label{tabstatic} Static properties associated with the $N(940)$ and $N(1535)$ elastic form factors, with $\kappa = F_2(0)$.
Where comparison is possible, results are consistent with those in Ref.\,\cite{Wilson:2011aa}.
($m_+ = 1.14\,$GeV is the nucleon dressed-quark core mass.)}
\begin{center}
\begin{tabular*}
{\hsize}
{
l|@{\extracolsep{0ptplus1fil}}
c@{\extracolsep{0ptplus1fil}}
c@{\extracolsep{0ptplus1fil}}
c@{\extracolsep{0ptplus1fil}}
c@{\extracolsep{0ptplus1fil}}}
\hline
& $N^+(1535)$ & $N^+(940)$ & $N^0(1535)$ & $N^0(940)$ \\
\hline
$r_1 \, m_+$ & $\phantom{-}3.20(22)$ & $3.34(15)$ & $0.91(35)$& $\phantom{-}0.88(34)$ \\
$r_2 \, m_+$ & $\phantom{-}3.52(76)$ & $3.46(62)$ & $3.39(36)$ & $\phantom{-}3.53(34)$ \\
$\kappa$ & $-1.18(46)$ & $1.36(34)$ & $0.68(29)$ & $-1.09(17)$ \\\hline
\end{tabular*}
\end{center}
\end{table}

It is worth noting that, with currents defined as in Eq.\,\eqref{JNNastExplicit} and for both the charged and neutral states, $F_1^-$ and $F_1^+$ have the same sign, but the sign of $F_2^-$ is opposite to that of $F_2^+$.

\section{Helicity amplitudes: $\gamma^{(\ast)} p \to N(1535)$ }
\label{sec:AAs}
As apparent in Eqs.\,\eqref{Belastic}, \eqref{N1535transition}, the natural focus for theoretical analyses of baryon elastic and transition form factors are the Dirac and Pauli form factors; and Sec.\,\ref{sec:FFs} presents our results in this form.  Experimental data on nucleon-to-resonance transitions, however, are usually presented in terms of helicity amplitudes (transverse $A_{1/2}$ and longitudinal $S_{1/2}$), which may be expressed in terms of $F_{1,2}^\ast$:
{\allowdisplaybreaks
\begin{subequations}
\begin{align}
A_{1/2} &= 2 {\cal K} \left( F_1^\ast + \frac{m_{N^\ast}-m_N}{m_{N^\ast}+m_N} F_2^\ast \right) \,, \\
S_{1/2} &= -\sqrt{2} {\cal K} (m_{N^\ast}+m_N) \frac{|\vec{q}\,|}{Q^2} \nonumber \\
& \quad \times \left( \frac{m_{N^\ast}-m_N}{m_{N^\ast}+m_N} F_1^\ast -\tau F_2^\ast \right)  \,,
\end{align}
where $\tau = Q^2/[m_{N^\ast}+m_N]^2$,
$|\vec{q}| = |Q_+| |Q_-|/[2 m_{N^\ast}]$,
\begin{align}
{\cal K}^2 &= \frac{\pi\alpha_{\rm em} |Q_{+}|^2}{2m_N(m_{N^\ast}^2-m_N^2)} \,,
\end{align}
\end{subequations}
with $\alpha_{\rm em}=e^2/[4\pi]$, the fine structure constant of quantum electrodynamics, and $|Q_{\pm}|^2 = (m_{N^\ast}\pm m_N)^2+Q^2$.  Hereafter, $m_N=m_+$ and $m_{N^\ast} = m_-$.
}

\begin{figure}[t]
\vspace*{1ex}

\rightline{\hspace*{0.5em}{\large{\textsf{A}}}}
\vspace*{-3ex}
\includegraphics[width=0.42\textwidth]{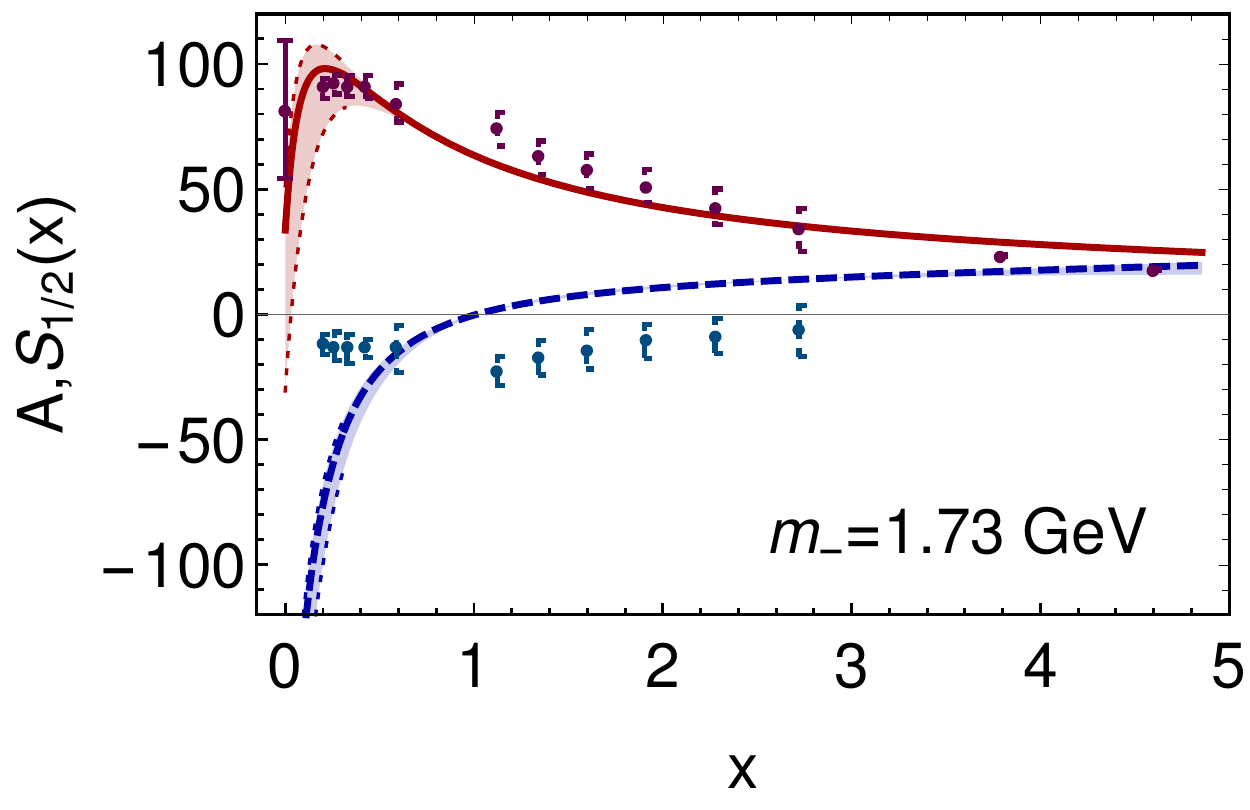}
\vspace*{1ex}

\rightline{\hspace*{0.5em}{\large{\textsf{B}}}}
\vspace*{-3ex}
\includegraphics[width=0.42\textwidth]{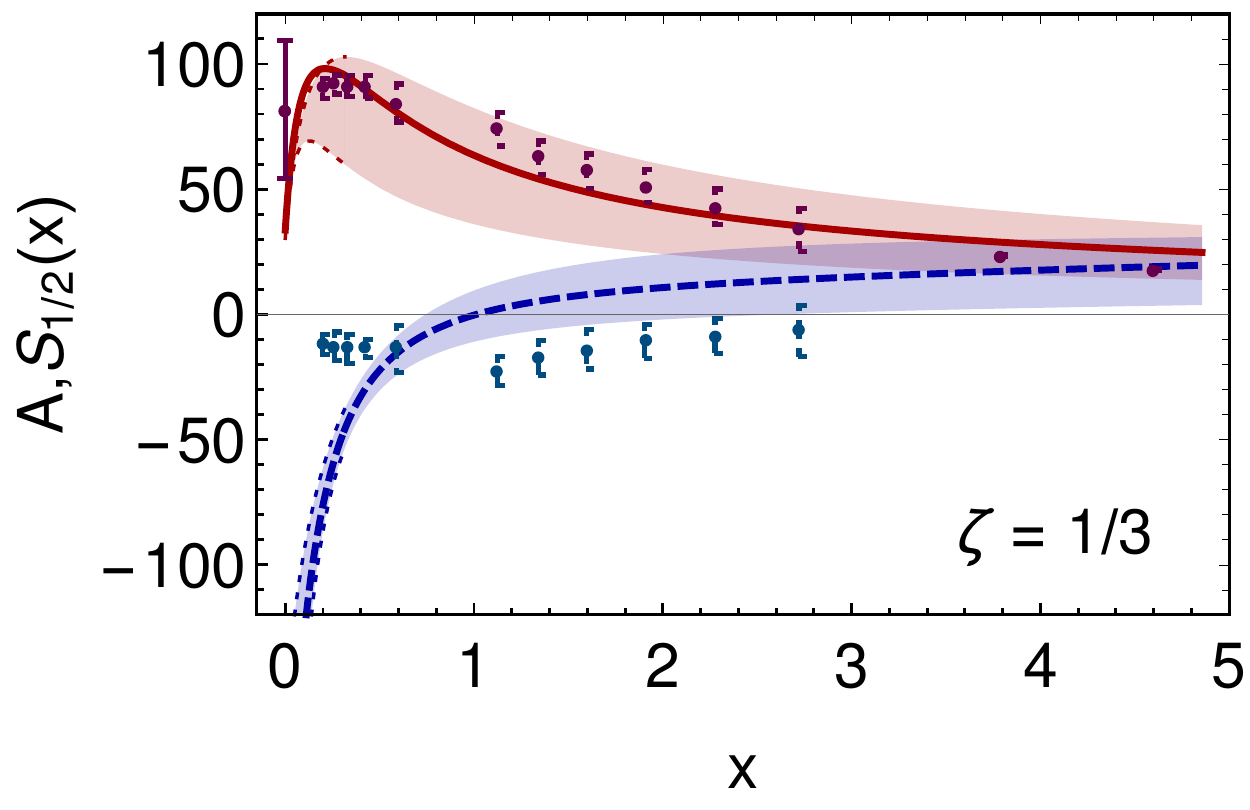}

\caption{
\emph{Upper panel}\,--\,{\sf A}.
$\gamma^{(\ast)} p\to N(1535)\frac{1}{2}^-$ helicity amplitudes as function of $x=Q^2/\bar{m}^2$, $\bar{m}=(m_++m_-)/2$: $A_{1/2}$ -- solid red; $S_{1/2}$ -- dashed blue.
Each central curve was obtained with the baryon masses in Eq.\,\eqref{BothMasses}, amplitudes in Eq.\,\eqref{eq:amplitudes}, and dressed-quark anomalous magnetic moment (DqAMM, \ref{SecS1}) $\zeta = 1/3$.  The associated shaded band indicates the response to variations of $\zeta \in [0.0,0.5]$: in both cases, $\zeta=0.5$ produces the uppermost curve.
\emph{Lower panel}\,--\,{\sf B}.
With $\zeta = 1/3$, response of helicity amplitudes to the variation $g_{DB}=0.2(1\pm0.5)$, Eq.\,\eqref{eq:amplitudesgDB}: smaller $g_{DB}$ produces the uppermost curve.
%
Experimental data are from Ref.\,\cite{Aznauryan:2009mx}.
\label{fig:HAsBoth}}
\end{figure}

\begin{figure}[t]
\vspace*{1ex}

\rightline{\hspace*{0.5em}{\large{\textsf{A}}}}
\vspace*{-3ex}
\includegraphics[width=0.42\textwidth]{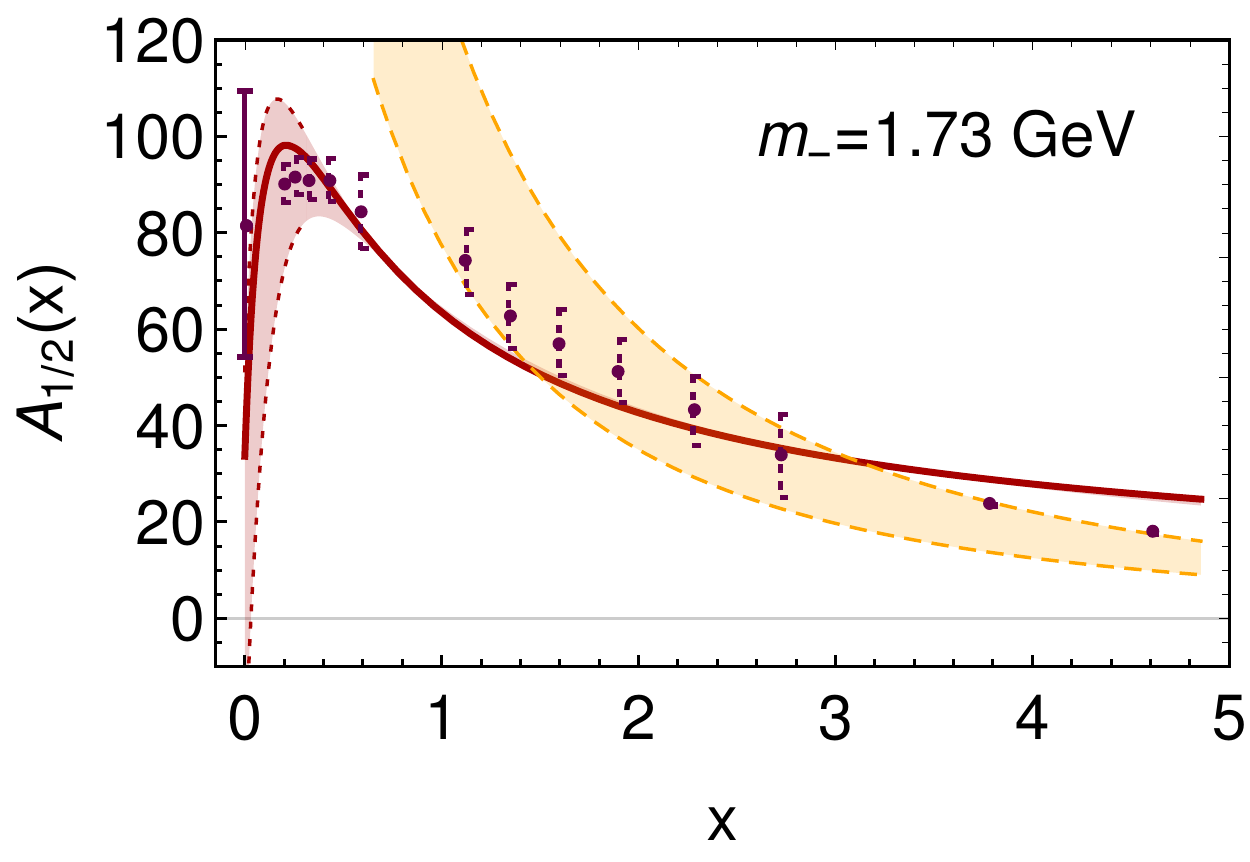}
\vspace*{1ex}

\rightline{\hspace*{0.5em}{\large{\textsf{B}}}}
\vspace*{-3ex}
\includegraphics[width=0.42\textwidth]{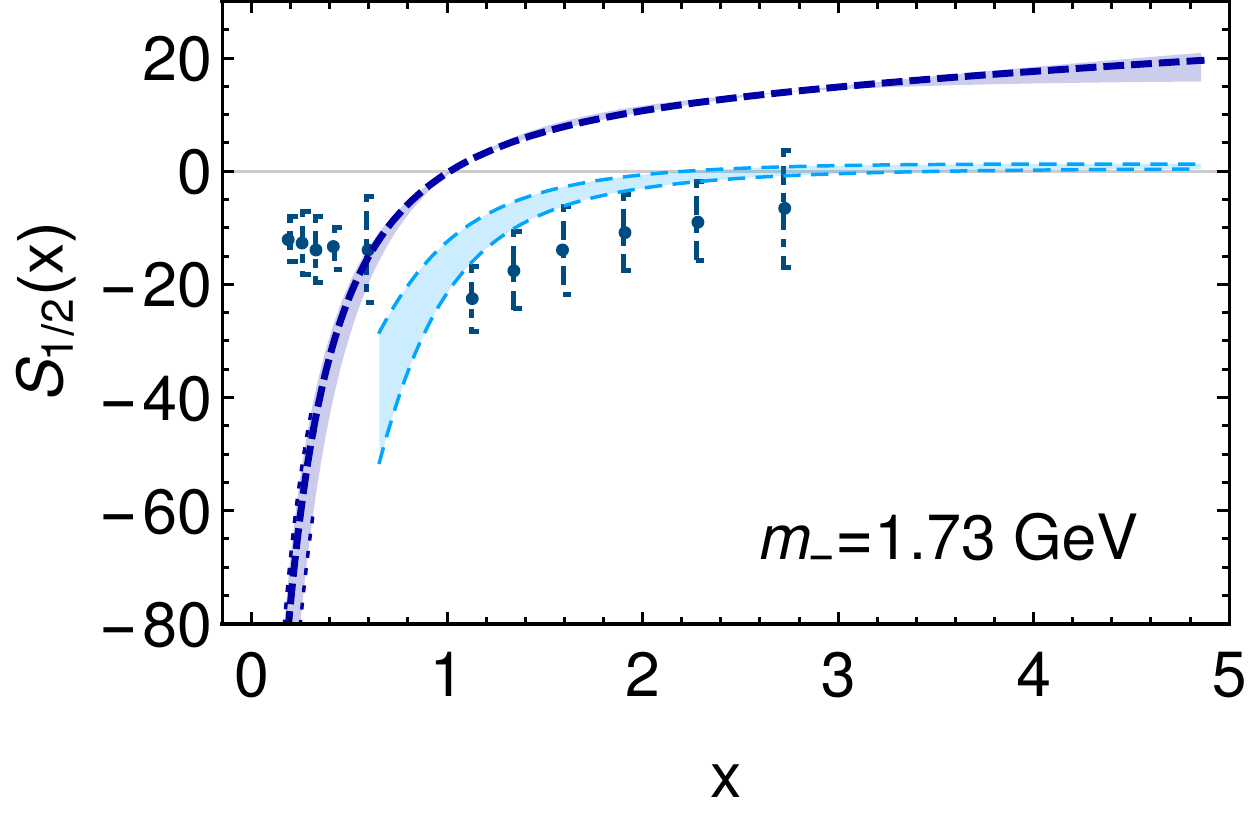}

\caption{
\emph{Upper panel}\,--\,{\sf A}.
$A_{1/2}$ for the $\gamma^{(\ast)} p\to N(1535)\,1/2^-$ transition, $x=Q^2/\bar{m}^2$,  $\bar{m}=(m_++m_-)/2$.  Solid red curve and DqAMM variation band -- SCI result; orange band bordered by dashed curves -- lattice-QCD (lQCD) assisted light-cone sum rules result \cite{Braun:2009jy1}.
\emph{Lower panel}\,--\,{\sf B}.
$S_{1/2}$: SCI result -- dashed blue curve within DqAMM variation band; and dashed light-blue curves with band -- Ref.\,\cite{Braun:2009jy1}.
Experimental data are from Ref.\,\cite{Aznauryan:2009mx}.
\label{fig:SA182}}
\end{figure}

We depict our SCI results for the transverse and longitudinal helicity amplitudes in Fig.\,\ref{fig:HAsBoth}.
Fig.\,\ref{fig:HAsBoth}\,A shows that the transverse amplitude is sensitive to the DqAMM; but with $\zeta \approx 1/3$, the SCI delivers a good description of modern data \cite{Aznauryan:2009mx}.  On the other hand, the longitudinal amplitude is practically insensitive to the DqAMM and there is no value for which the SCI delivers a quantitatively good description of the data.  As will be seen in Sec.\,\ref{sec:FFs}, this is because the SCI result for $F_2^\ast$ is hard, \emph{viz}.\ it falls too slowly with increasing $x$.

\begin{figure}[t]
\vspace*{1ex}

\rightline{\hspace*{0.5em}{\large{\textsf{A}}}}
\vspace*{-3ex}
\includegraphics[width=0.42\textwidth]{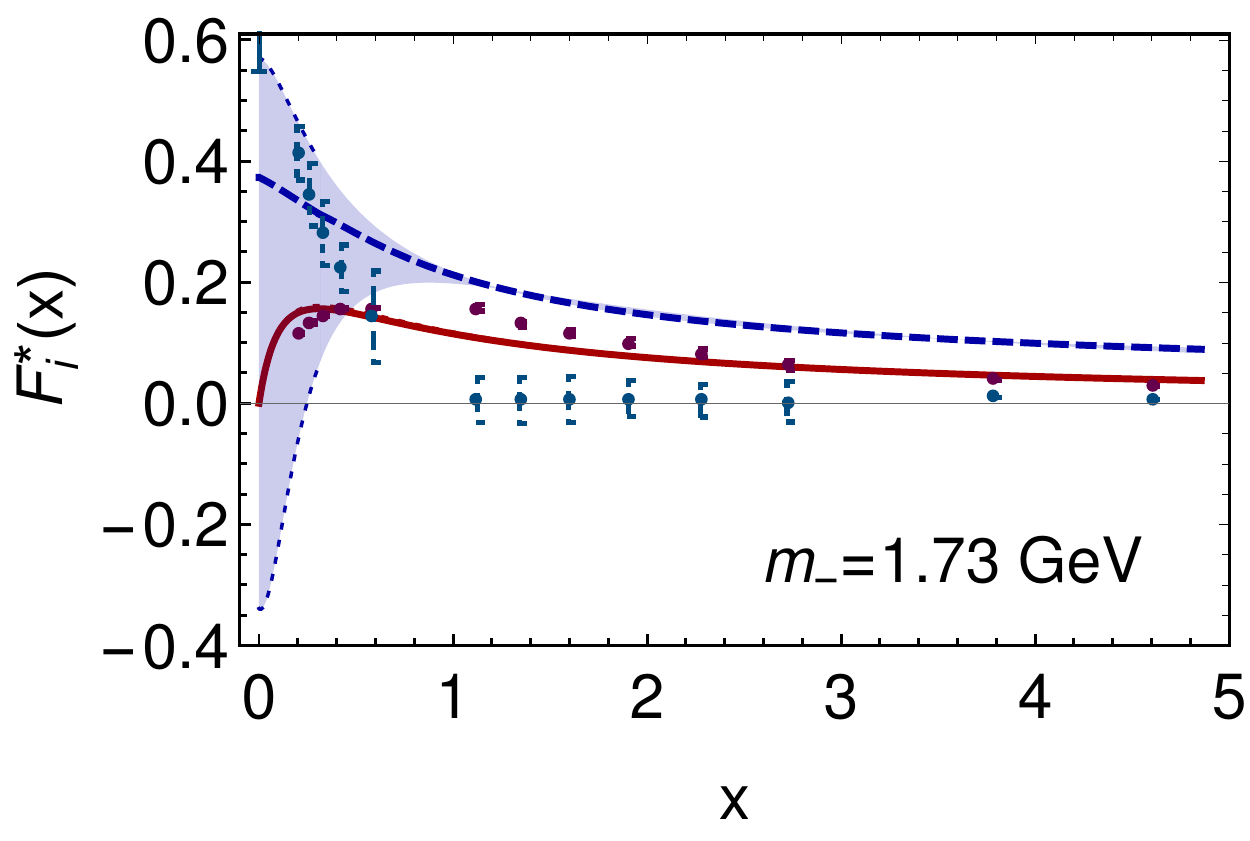}
\vspace*{1ex}

\rightline{\hspace*{0.5em}{\large{\textsf{B}}}}
\vspace*{-3ex}
\includegraphics[width=0.42\textwidth]{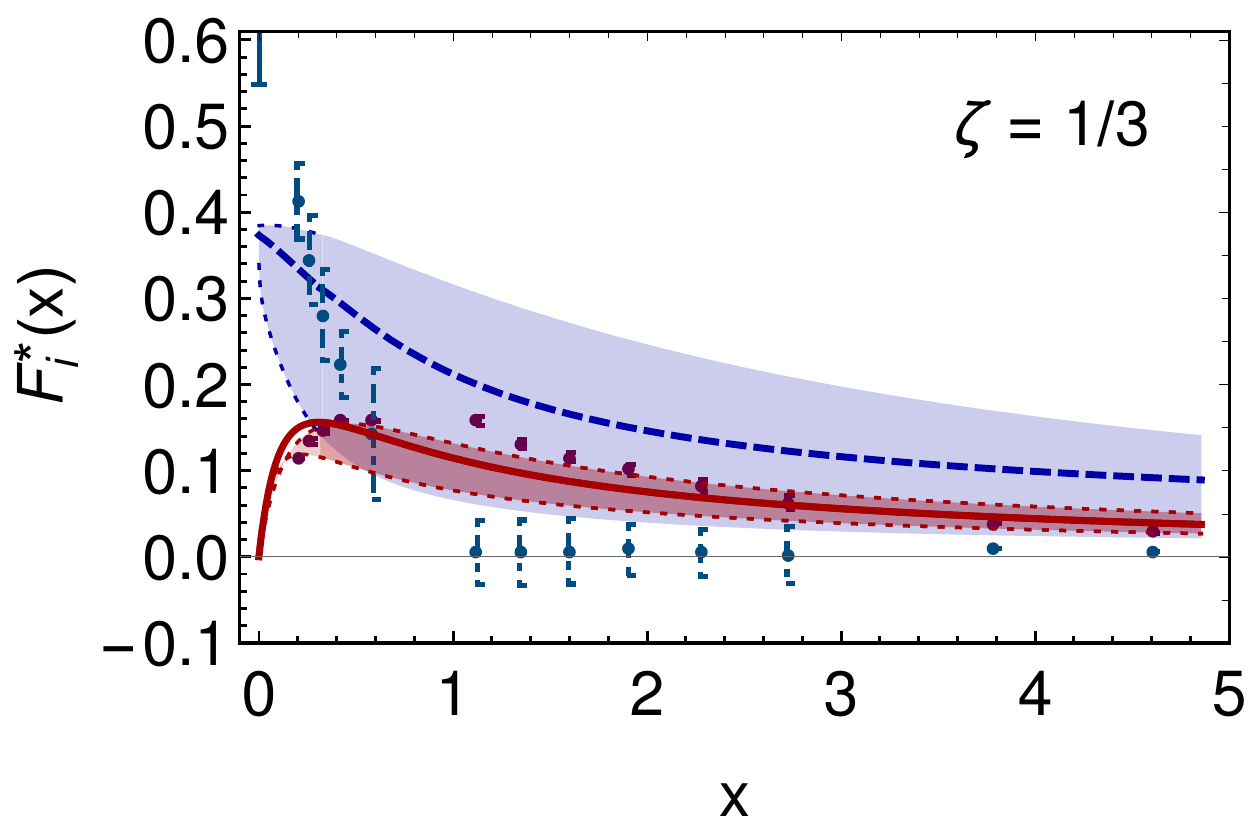}

\caption{
\emph{Upper panel}\,--\,{\sf A}.
$\gamma^{(\ast)} p\to N(1535)\frac{1}{2}^-$ Dirac and Pauli transition form factors as function of $x=Q^2/\bar{m}^2$, $\bar{m}=(m_++m_-)/2$: solid red -- $F_1^\ast$; dashed blue -- $F_2^\ast$.
Each central curve was obtained with the baryon masses in Eq.\,\eqref{BothMasses}, amplitudes in Eq.\,\eqref{eq:amplitudes}, and DqAMM $\zeta = 1/3$.  The associated shaded band (invisible for $F_1^\ast$) indicates the response to variations of $\zeta \in [0.0,0.5]$: $\zeta=0.5$ produces the uppermost curve.
\emph{Lower panel}\,--\,{\sf B}.
With $\zeta = 1/3$, response of transition form factors to the variation $g_{DB}=0.2(1\pm0.5)$, Eq.\,\eqref{eq:amplitudesgDB}: smaller $g_{DB}$ produces the uppermost curve.
%
Experimental data are reconstructed from Ref.\,\cite{Aznauryan:2009mx}.
\label{fig:FFsBoth}}
\end{figure}

Fig.\,\ref{fig:HAsBoth}\,B displays the sensitivity of the SCI helicity amplitudes to changes $g_{DB}=0.2(1\pm0.5)$.  This parameter is included in the Faddeev kernel in order to model the impact of DCSB-enhanced ``spin-orbit'' repulsion effects, which are typically underestimated in RL truncation \cite{Qin:2020jig}.   As discussed in connection with Eq.\,\eqref{eq:amplitudesgDB}, this variation shifts the quark-core mass of the $N(1535)\,1/2^-$  and, more importantly, it changes the character of the Faddeev amplitude -- a larger value of $[1-g_{DB}]$ means more repulsion; so, a higher fraction of negative-parity diquarks and a larger core mass.  Both helicity amplitudes are sensitive to $g_{DB}$-induced changes in the $N(1535)\,1/2^-$ wave function.  This emphasises the importance of such resonance electrocouplings: with nucleon structure well constrained, they are keen probes of the structure of the final state baryon, possessing far greater sensitivity than that baryon's mass alone.
In the SCI context, the best description of data is obtained when the scalar and pseudoscalar diquark content of the $N(1535)\,\tfrac{1}{2}^-$ are balanced.

Fig.\,\ref{fig:SA182} compares SCI results for the helicity amplitudes with those obtained using lQCD input to inform a light-cone sum rules calculation \cite{Braun:2009jy1}.  Evidently, the two approaches agree on the magnitude of $A_{1/2}$ and the low-$x$ sign of $S_{1/2}$.  The sign-change in the SCI result for $S_{1/2}$ is again an artefact of the hardness of $F_2^\ast$.

\section{Form factors}
\label{sec:FFs}
Our predictions for the $\gamma^{(\ast)} p \to N(1535)\,\tfrac{1}{2}^-$ Dirac and Pauli transition form factors are drawn in Fig.\,\ref{fig:FFsBoth}.
Regarding Fig.\,\ref{fig:FFsBoth}\,A, the SCI result for $F_1^\ast$ is insensitive to the DqAMM and in fair quantitative agreement with data.  On the other hand, while that obtained with $\zeta=1/3$ for $F_2^\ast$ agrees in magnitude with data on $x\lesssim 1$, $F_2^\ast$ is very sensitive to the DqAMM on this domain and too hard on its complement.  As will be seen in Sec.\,\ref{sec:Dissections}, this latter feature owes to the fact that, on $x\gtrsim 1$, $F_2^\ast$ is dominated by Diagram S1 in Fig.\,\ref{fig:current} -- photon scattering from quark; and the absence of momentum dependence in the SCI Faddeev amplitude entails that this diagram is (unphysically) hard.

\begin{figure}[t]
\vspace*{1ex}

\rightline{\hspace*{0.5em}{\large{\textsf{A}}}}
\vspace*{-3ex}
\includegraphics[width=0.42\textwidth]{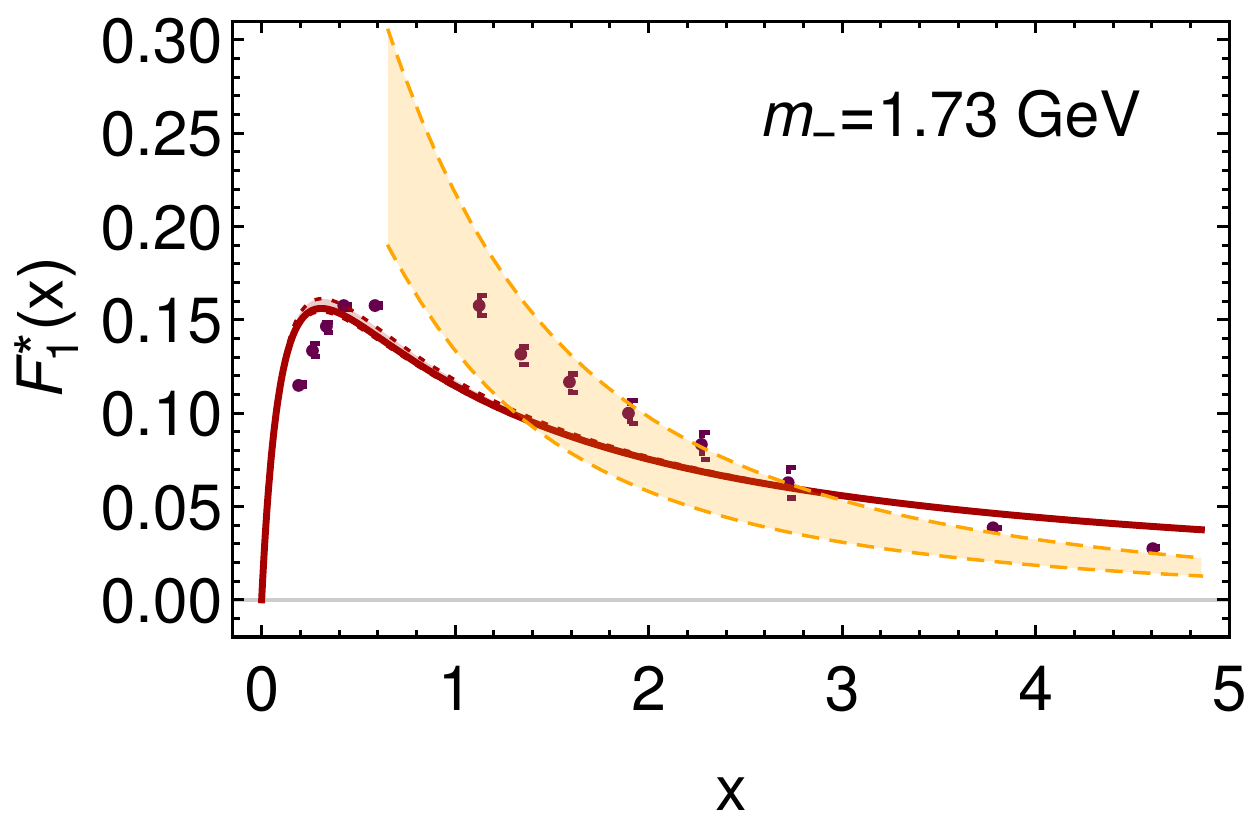}
\vspace*{1ex}

\rightline{\hspace*{0.5em}{\large{\textsf{B}}}}
\vspace*{-3ex}
\includegraphics[width=0.42\textwidth]{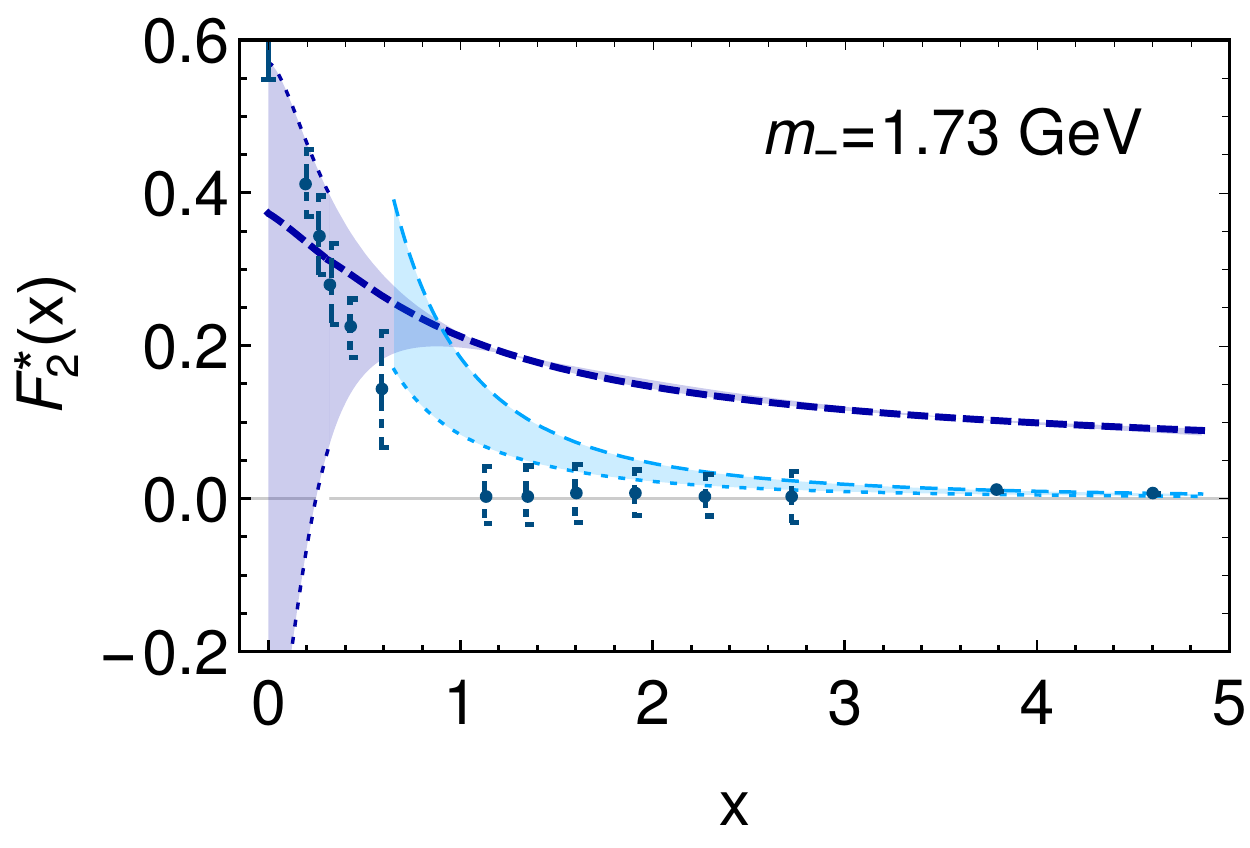}

\caption{
\emph{Upper panel}\,--\,{\sf A}.
$\gamma^{(\ast)} p\to N(1535)\frac{1}{2}^-$ Dirac transition form factor as function of $x=Q^2/\bar{m}^2$, $\bar{m}=(m_++m_-)/2$.  Solid red curve and DqAMM variation band -- SCI result; orange band bordered by dashed curves -- lQCD-assisted light-cone sum rules result \cite{Braun:2009jy1}.
\emph{Lower panel}\,--\,{\sf B}.
Analogous Pauli transition form factor: SCI result -- dashed blue curve within DqAMM variation band; and dashed light-blue curves with band -- Ref.\,\cite{Braun:2009jy1}.
Experimental data reconstructed from Ref.\,\cite{Aznauryan:2009mx}.
\label{fig:F1F2182}}
\centering
\end{figure}

Fig.\,\ref{fig:FFsBoth}\,B reveals the sensitivity of the Dirac and Pauli transition form factors to changes in the internal structure of the $N(1535)\,\tfrac{1}{2}^-$.  The reaction of $F_1^\ast$ is modest, but $F_2^\ast$ responds strongly.  Again, this owes to the dominance of Diagram S1 in Fig.\,\ref{fig:current} -- scalar diquarks do not have magnetic interactions -- and interference in that diagram between $N(1535)\,\tfrac{1}{2}^-$ scalar-diquark strength and compensating resonance-mass effects.

Fig.\,\ref{fig:F1F2182} compares SCI predictions for the transition form factors with available lQCD-assisted light-cone sum rules results \cite{Braun:2009jy1}.  Here, on the domain of quark core dominance, there is fair agreement on $F_1^\ast$, but the SCI's limitations in connection with magnetic interactions is again evident in the $F_2^\ast$ mismatch.

\section{Form factor dissection}
\label{sec:Dissections}
As apparent in Fig.\,\ref{fig:current}, the $\gamma^{(\ast)} p\to N(1535)\frac{1}{2}^-$ transition current can be considered as a sum of three distinguishable terms.
\begin{itemize}
\item {\sf S1}.  Photon strikes a dressed-quark with an associated spectator diquark.  In this case, owing to the structure of the nucleon Faddeev amplitude and the discussion in \ref{Appgdq}, there is only one possible contribution, \emph{viz}.\ $0^+$ spectator diquark.
\item {\sf S2}.  The dressed-quark is a spectator to a photon+di\-quark elastic scattering event.  It follows from the absence of pseudoscalar and vector diquarks in the nucleon Faddeev amplitude that there are only two such contributions to the transition: $\gamma 0^+ \to 0^+$ and $\gamma 1^+ \to 1^+$.
\item {\sf S3}.  The dressed-quark is a spectator to a photon-induced diquark transition:
    $\gamma 0^+ \leftrightarrow 1^+$;
    $\gamma 0^+ \to  1^-$;
    $\gamma 1^+ \to 1^-$.
\end{itemize}
Evidently, there are six contributions in total.  We label them as follows:
\begin{itemize}
\item $Q^+Q^+ := ${\sf S1}.
\item $D^+ D^+:=$ {\sf S2}$\,+\, \gamma 0^+ \leftrightarrow 1^+$, i.e.\ the sum of all dressed-quark spectator terms with a positive-parity diquark in the initial and final states.
\item $D^- D^+:=\,$sum of the remaining two dressed-quark spectator terms with a positive-parity diquark in the initial state and a negative parity correlation in the final state.
\end{itemize}

\begin{figure}[t]
\vspace*{1ex}

\rightline{\hspace*{0.5em}{\large{\textsf{A}}}}
\vspace*{-3ex}
\includegraphics[width=0.42\textwidth]{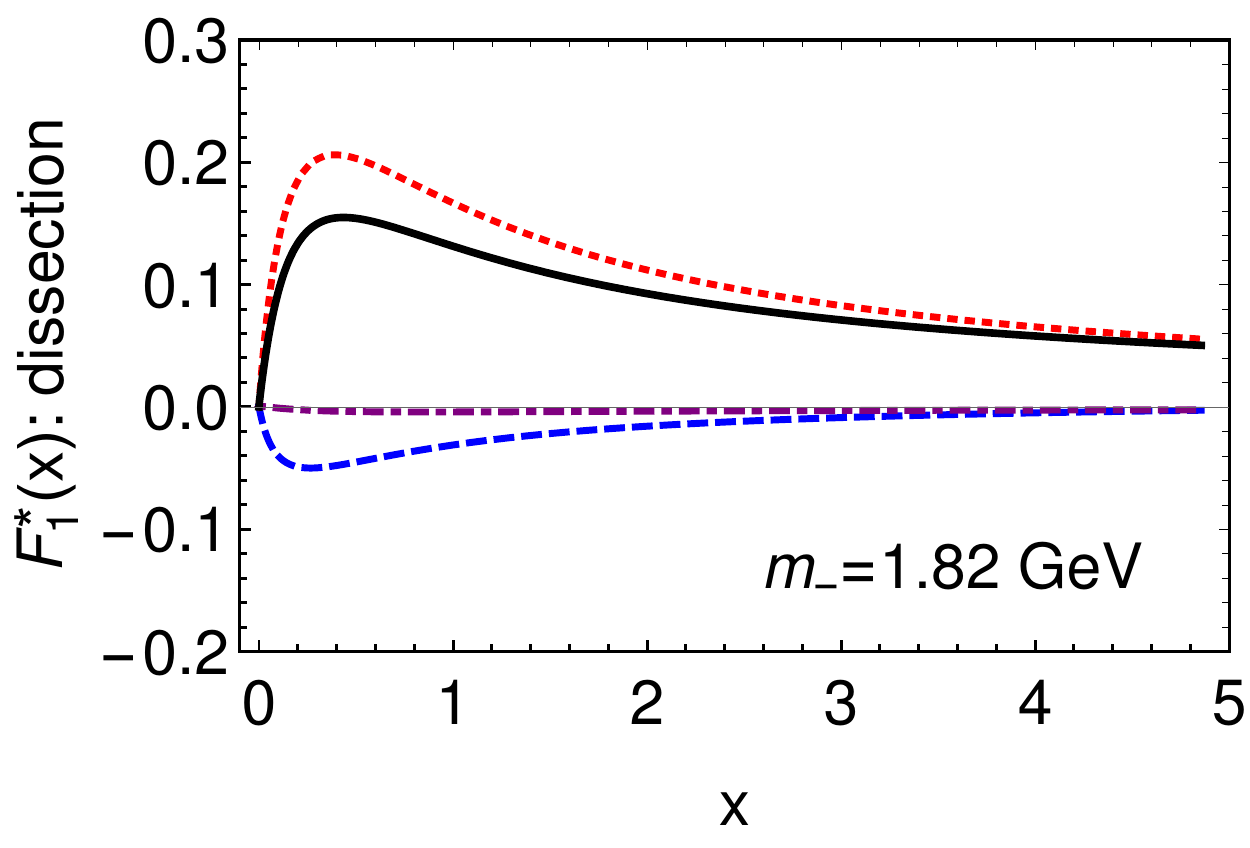}
\vspace*{1ex}

\rightline{\hspace*{0.5em}{\large{\textsf{B}}}}
\vspace*{-3ex}
\includegraphics[width=0.42\textwidth]{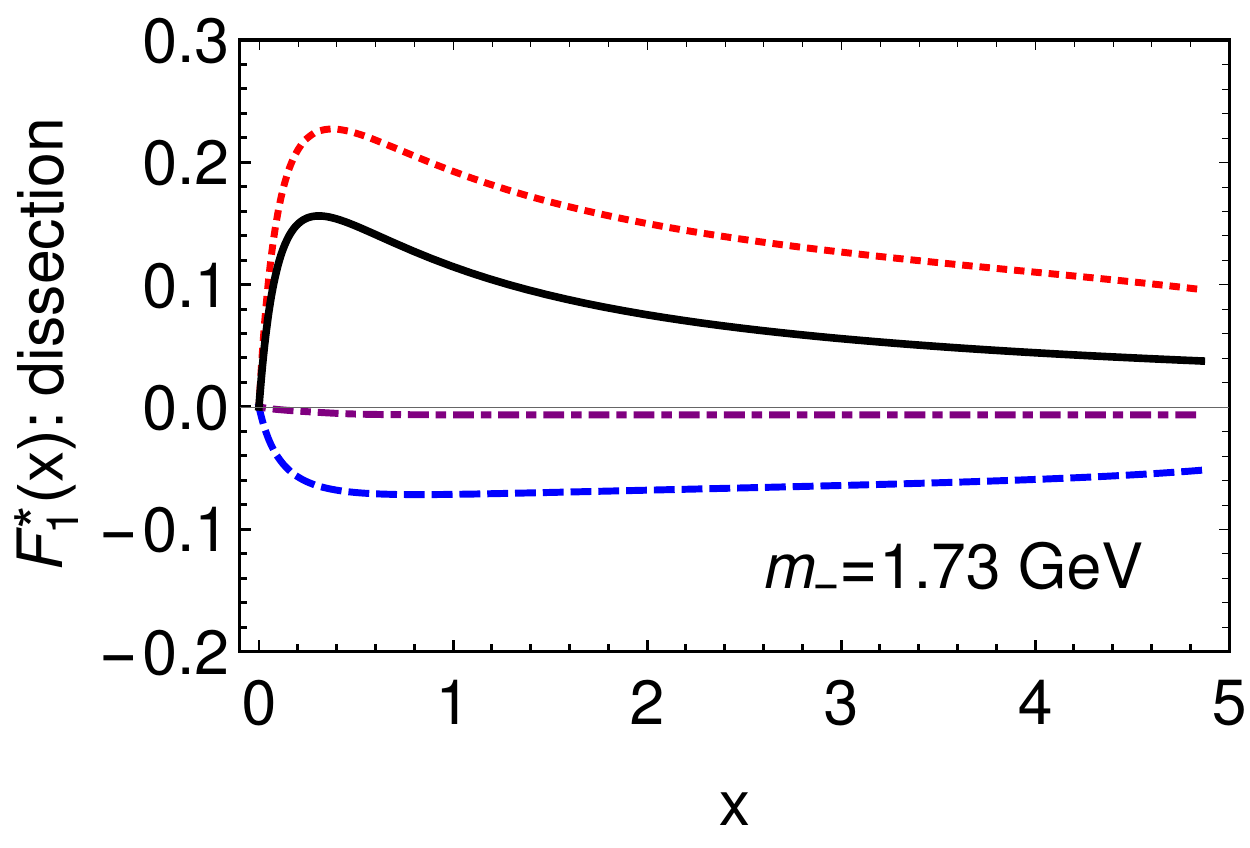}

\rightline{\hspace*{0.5em}{\large{\textsf{C}}}}
\vspace*{-3ex}
\includegraphics[width=0.42\textwidth]{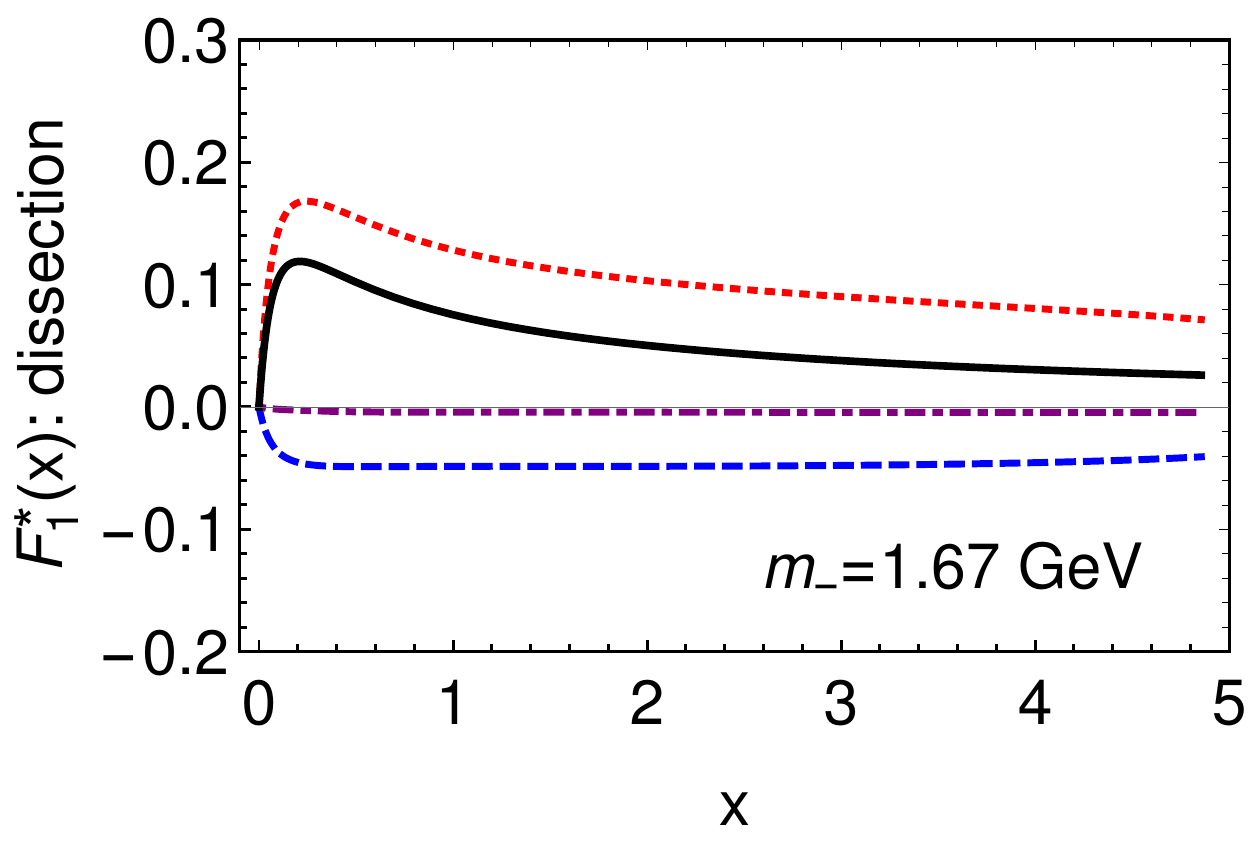}

\caption{
$\gamma^{(\ast)} p \to N(1535)\frac{1}{2}^-$ Dirac transition form factor computed with $\zeta = 1/3$.  As usual, $x=Q^2/\bar{m}^2$, with $\bar{m}=(m_++m_-)/2$.
Black solid curve -- complete result.
Dissection:
$Q^+Q^+$ -- dotted red;
$D^+D^+$ -- dashed blue;
$D^-D^+$ -- dot-dashed purple.
The three panels reveal the sensitivity to the mass and structure of the $N(1535)\frac{1}{2}^-$ final state.
\label{fig:DF1All}}
\end{figure}

In terms of the dissection just described, the $\gamma^{(\ast)} p \to N(1535)\frac{1}{2}^-$ Dirac transition form factor is drawn in Fig.\,\ref{fig:DF1All}.  Plainly, $D^-D^+$ contributions are negligible and the complete result is typically obtained from destructive interference between $Q^+ Q^+$ and $D^+D^+$.
The $m_- = 1.82\,$GeV result is somewhat special because, reading from Eq.\,\eqref{eq:amplitudesgDB}, in this case the $N(1535)\frac{1}{2}^-$ contains a little $0^+$ diquark and practically no $1^+$ diquark, so the $D^+D^+$ contribution is dominated by the $\gamma 0^+ \to 1^+$ transition, which is large at $x=0$ but vanishes with increasing $x$ -- see Fig.\,\ref{fig:1pm0pmDFF}.

\begin{figure}[t]

\vspace*{1ex}

\rightline{\hspace*{0.5em}{\large{\textsf{A}}}}
\vspace*{-3ex}
\includegraphics[width=0.42\textwidth]{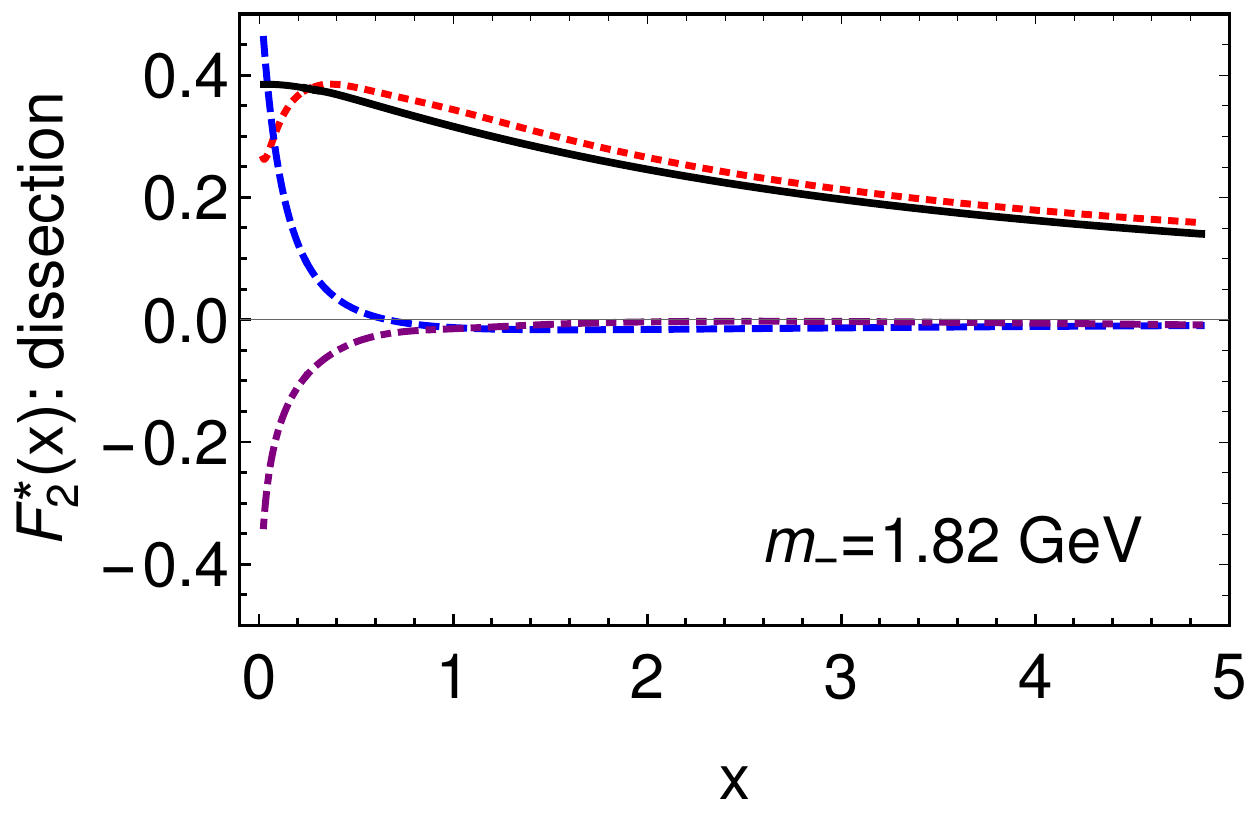}
\vspace*{1ex}

\rightline{\hspace*{0.5em}{\large{\textsf{B}}}}
\vspace*{-3ex}
\includegraphics[width=0.42\textwidth]{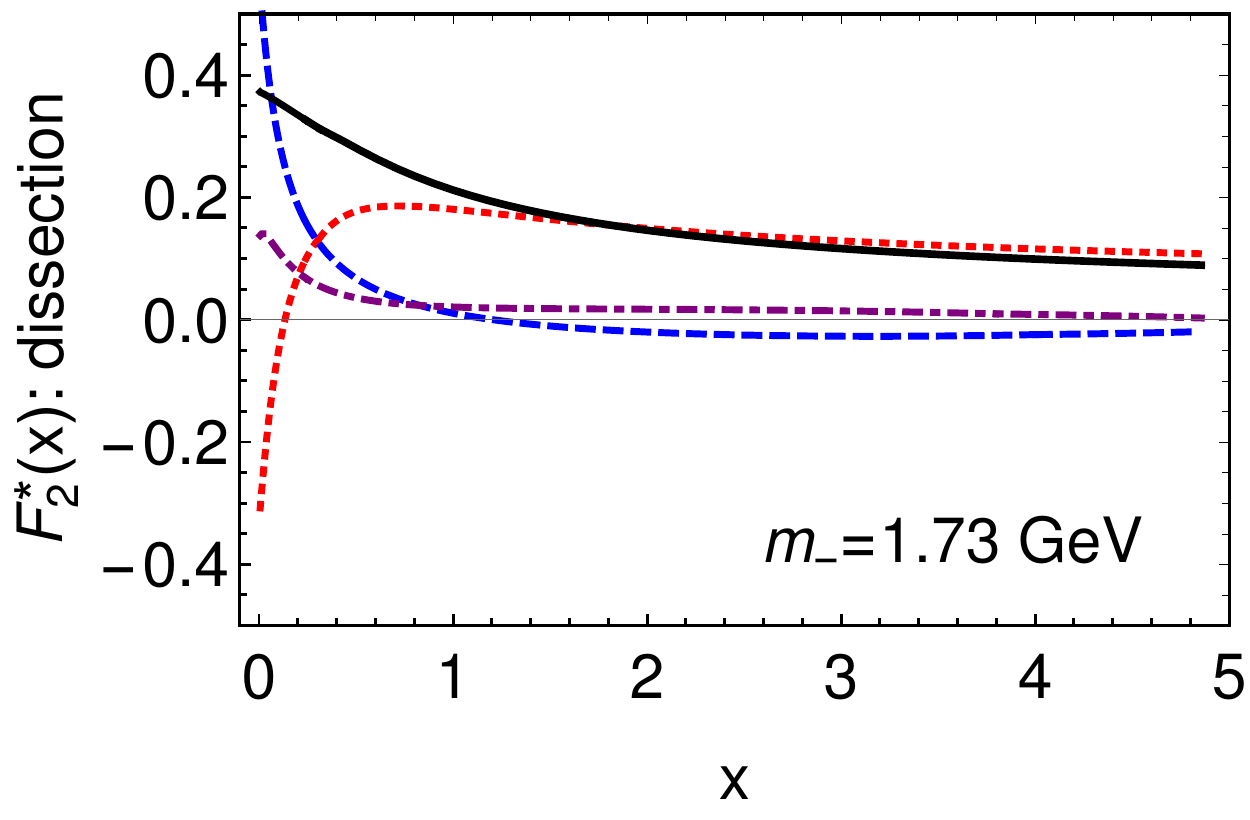}

\rightline{\hspace*{0.5em}{\large{\textsf{C}}}}
\vspace*{-3ex}
\includegraphics[width=0.42\textwidth]{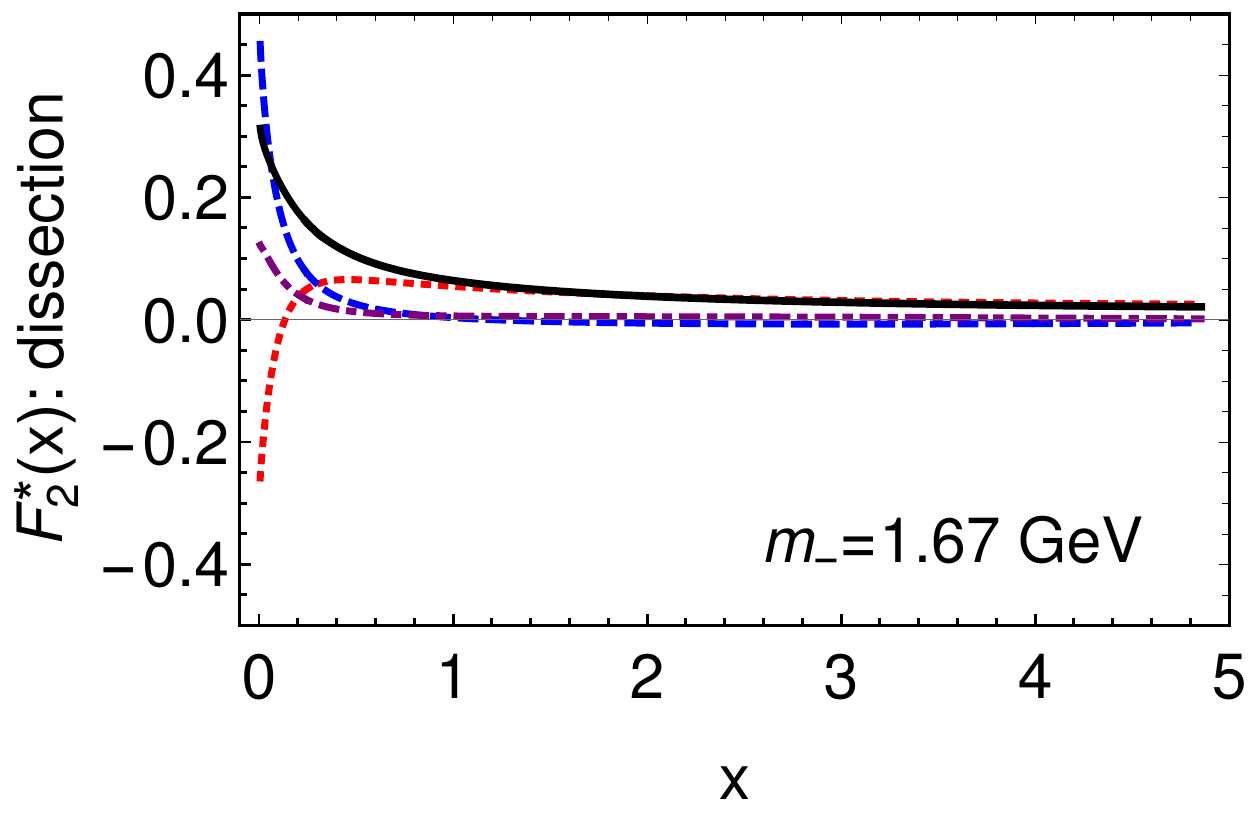}

\caption{
$\gamma^{(\ast)} p \to N(1535)\frac{1}{2}^-$ Pauli transition form factor computed with $\zeta = 1/3$.  As usual, $x=Q^2/\bar{m}^2$, with $\bar{m}=(m_++m_-)/2$.
Black solid curve -- complete result.
Dissection:
$Q^+Q^+$ -- dotted red;
$D^+D^+$ -- dashed blue;
$D^-D^+$ -- dot-dashed purple.
The three panels reveal the sensitivity to the mass and structure of the $N(1535)\frac{1}{2}^-$ final state.
\label{fig:DF2All}}
\end{figure}

Analogous results for the $\gamma^{(\ast)} p \to N(1535)\frac{1}{2}^-$ Pauli transition form factor are drawn in Fig.\,\ref{fig:DF2All}.
State orthogonality does not require the Pauli form factor to vanish at $x=0$, so the sensitivity of this magnetic form factor to $N(1535)\frac{1}{2}^-$ structure is marked on $x\lesssim 1$:
$D^+ D^+$ and $D^- D^+$ magnetic transitions interfere constructively for the lighter masses and destructively for the heaviest $N(1535)\frac{1}{2}^-$ mass;
and the low-$x$ behaviour of the $Q^+Q^+$ diagram expresses the impact of the DqAMM.  (Recall that the DqAMM strength drops rapidly with increasing $x$, Eq.\,\eqref{DQAMM}.)
On $x\gtrsim 1$, only the $Q^+ Q^+$ magnetic contribution survives.  This diagram is hard, with a value determined ultimately by competition between the decreasing scalar diquark content of the $N(1535)\frac{1}{2}^-$ and an enhancement, driven by its increasing mass, $m_-$, in the $F_2^\ast$ integrand.

\section{Summary and Perspective}
\label{Epilogue}
Using a confining, symmetry-preserving regularisation of a vector$\,\otimes\,$vector contact interaction (SCI) [Sec.\,\ref{SecSCI}], we computed all helicity amplitudes and form factors associated with the $\gamma^{(\ast)}+p \to N(1535) \tfrac{1}{2}^-$ transition on the entire domain of accessible momentum transfers.

In this analysis, the initial and final state baryons are described as quark+diquark bound-states obtained as dynamical solutions of a Poincar\'e-covariant Faddeev equation.  The nucleon solution is dominated by scalar and axial-vector diquark correlations, but its parity partner, the $N(1535) \tfrac{1}{2}^-$, also contains sizeable pseudoscalar and vector diquarks components [Sec.\,\ref{subsec:BoundState}].
The significance of these additional correlations required that we also compute a large array of hitherto unknown photon+diquark form factors, \emph{viz}.\ those involving negative-parity diquarks [\ref{ES2}, \ref{ES3}].

We found that both the Dirac and Pauli $\gamma^{(\ast)}+p \to N(1535) \tfrac{1}{2}^-$ transition form factors are sensitive to the structure of the Faddeev amplitudes of the baryons involved, especially that of the $N(1535) \tfrac{1}{2}^-$ [Sec.\,\ref{sec:FFs}].
$F_1^\ast$ is dominated by diagrams that contain a positive parity diquark in the initial and final state;
whereas $F_2^\ast$ is sensitive to interference between photon interactions with positive and negative parity  diquarks in the final state [Sec.\,\ref{sec:Dissections}].
The (magnetic) Pauli transition form factor is also sensitive to the presence of a dressed-quark anomalous magnetic moment and this is expressed strongly in the transverse helicity amplitude, $A_{1/2}$ [Sec.\,\ref{sec:AAs}].

Overall, the SCI provides a fair description of existing data.  Naturally, SCI studies have limitations, primarily evident in form factor $Q^2$-dependence that is too hard on $Q^2 \gtrsim m_N^2$, where $m_N$ is the nucleon mass.  On the other hand, they have the merit of being largely algebraic.  Hence, SCI analyses are capable of quickly delivering valuable insights.  Experience shows that these qualitative outcomes translate into similar effects in frameworks built upon elements with closer connections to QCD, e.g.: dressed-quarks with momentum-dependent masses; diquarks with momentum-dependent correlations amplitudes; and, consequently, baryons described by sophisticated, momentum-dependent Faddeev amplitudes.  An analysis of the $\gamma^{(\ast)}+p \to N(1535) \tfrac{1}{2}^-$ transition in such an approach is underway.

\begin{acknowledgements}
%
%
We are grateful for: input from D.\,J.~Wilson in the early stages of this project; 
and constructive comments from
V.~Mokeev.
%
%
%
Work supported by:
National Natural Science Foundation of China (under Grant No.\,11805097);
Jiangsu Provincial Natural Science Foundation of China (under Grant No.\,BK20180323);
Coordinaci\'on de la Investigaci\'on Científica (CIC) of the University of Michoacan and CONACyT, Mexico, through grant nos.\ 4.10 and CB2014-22117, respectively;
%
%
Ministerio Espa\~nol de Ciencia e Innovaci\'on, grant no. PID2019-107844GB-C22;
and Junta de Andaluc\'ia, contract nos.\ P18-FR-5057 and Operativo FEDER Andaluc\'ia 2014-2020 UHU-1264517.
\end{acknowledgements}

\appendix
\section{Electromagnetic interaction vertices}
\label{App:current}
To calculate the baryon elastic and transition currents considered herein, the vertices in Fig.\,\ref{fig:current} must be specified, \emph{i.e}.\ the momentum-dependent photon+quark and photon+diquark interaction form factors.

\subsection{Photon+quark vertex -- S1}
\label{SecS1}
\begin{figure}[!t]
\includegraphics[width=0.42\textwidth]{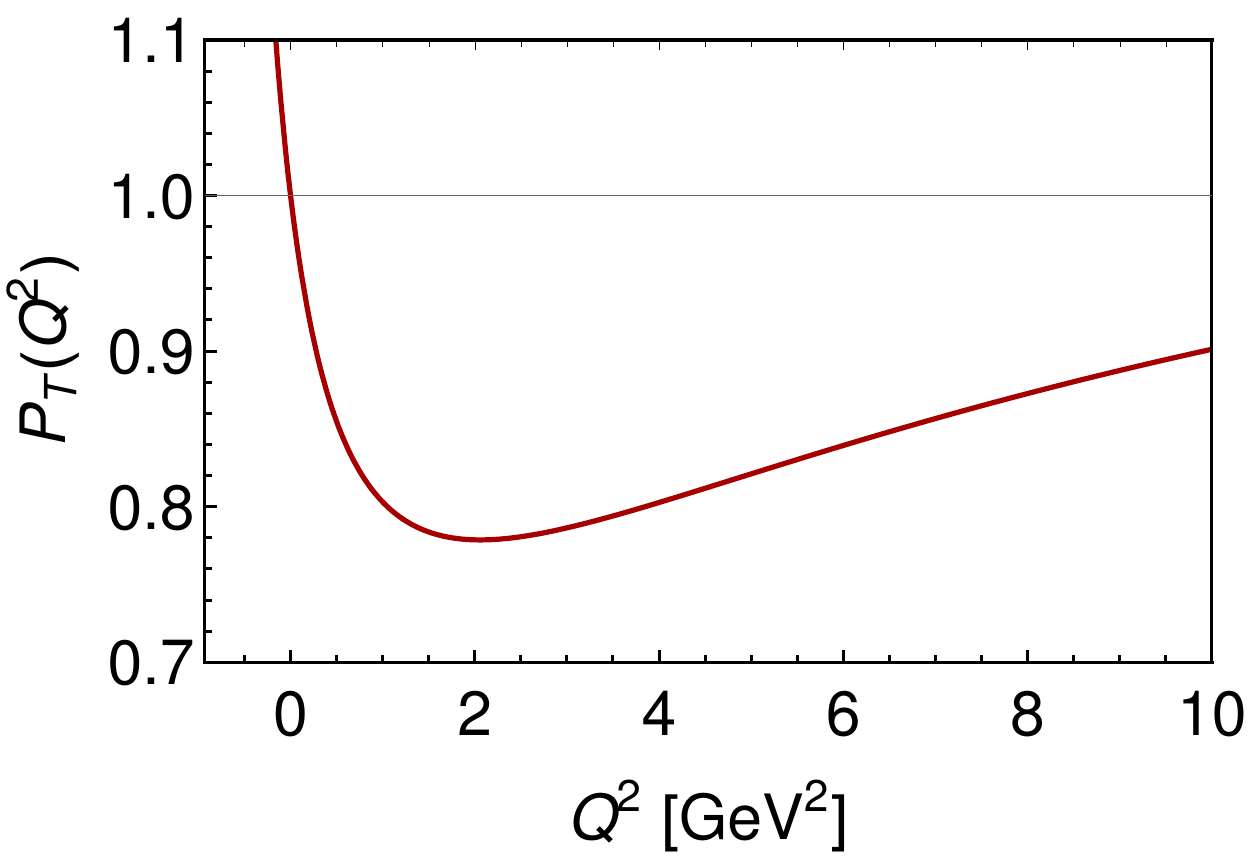}
\caption{Photon+quark vertex dressing function in Eq.\,\eqref{EqPT}.  As in any symmetry preserving treatment of photon+quark interactions, $P_{\rm T}(Q^2)$ exhibits a pole at $Q^2= -m_\rho^2$.  Moreover, $P_{\rm T}(Q^2 = 0) = 1 = P_{\rm T}(Q^2 \to \infty)$.
\label{fig:rhoPole}}
\end{figure}

The primary element throughout is the dressed photon+quark vertex, which takes the following form when using the SCI:
\begin{equation}
\Gamma_\mu^\gamma(Q) = \frac{Q_\mu Q_\nu}{Q^2} \gamma_\nu + \Gamma_\mu^{\rm T}(Q)\,,
\end{equation}
with
$Q=p_f-p_i$, where $p_{f,i}$ are the outgoing, incoming quark momenta,
$\Gamma_\mu^{\rm T}(Q) = {\mathpzc P}_{\mu\nu}(Q) \Gamma_\nu(Q)$,
${\mathpzc P}_{\mu\nu}(Q) = \delta_{\mu\nu} - Q_\mu Q_\nu / Q^2$,
and \cite{Roberts:2010rn, Wilson:2011aa}:
\begin{align}
\Gamma_\mu^{\rm T}(Q) &= P_{\rm T}(Q^2) {\mathpzc P}_{\mu\nu}(Q) \gamma_\nu \nonumber \\
&
\quad + \frac{\zeta}{2 M_u} \sigma_{\mu\nu} Q_\nu \exp\left(-\frac{Q^2}{4 M_u^2}\right) \,.
\label{DQAMM}
\end{align}
Here \cite{Roberts:2010rn}
\begin{subequations}
\label{EqPT}
\begin{align}
P_{\rm T}(Q^2) & = \frac{1}{1+K_\gamma(Q^2)}\,, \\
K_\gamma(Q^2) & = \frac{4\alpha_{\rm IR} Q^2}{3\pi m_G^2}
\int_0^1 d\alpha\,\alpha(1-\alpha)\,\bar{\cal C}_1(\omega(\alpha,\Delta^2))\,,
\end{align}
\end{subequations}
where
the mass-scale $m_G=0.5\,$GeV when $\alpha_{IR}$ has the value in Table~\ref{tabledressedquark},
$\omega(\alpha,Q^2) = M^2 + \alpha (1-\alpha) Q^2$,
\begin{align}
\overline{\cal C}_1(\sigma) & = \Gamma(0,\sigma \tau_{\rm ir}^2) - \Gamma(0,\sigma \tau_{\rm uv}^2)\,,
\end{align}
with $\Gamma(\alpha,y)$ being the incomplete gamma-function.  The dressing function in Eq.\,\eqref{EqPT} is depicted in Fig.\,\ref{fig:rhoPole}.

The second term in Eq.\,\eqref{DQAMM} expresses the fact that owing to DCSB a dressed light-quark has a large anomalous electromagnetic moment (DqAMM) \cite{Chang:2010hb, Xing:2021dwe}. With $\zeta=1/3$, we reproduce all form factor results in Ref.\,\cite{Wilson:2011aa}.  Our value for $\zeta$ is smaller than that used therein because Ref.\,\cite{Wilson:2011aa} omitted this contribution when computing the $0^+ \leftrightarrow 1^+$ diquark transition form factor.  To illustrate the sensitivity of calculated observables to the DqAMM, we typically show results obtained with $\zeta \in [0,0.5]$, highlighting those obtained using $\zeta=1/3$.

\subsection{Elastic photon+diquark vertices -- S2}
\label{ES2}
\begin{figure}[!t]
\includegraphics[width=0.42\textwidth]{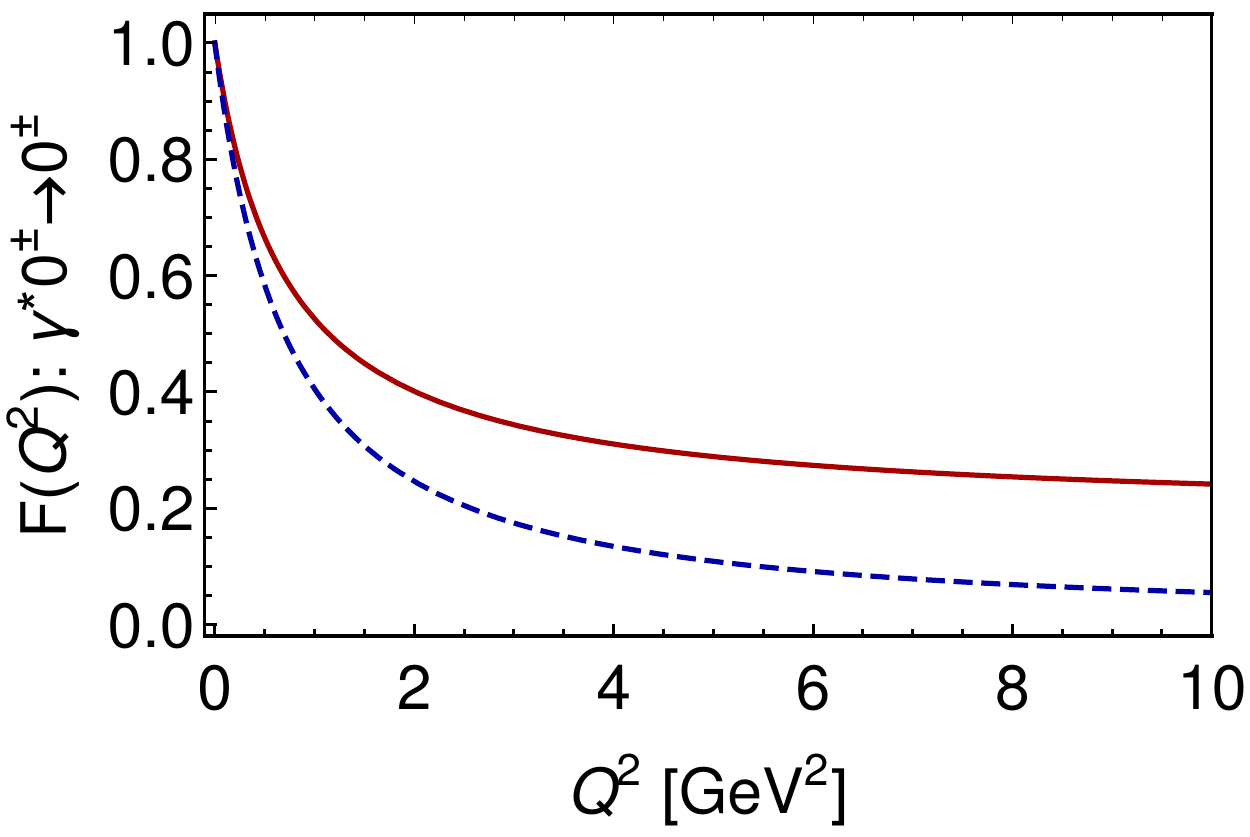}
\caption{\label{fig:0pm0pmDFF}
Elastic photon+(pseudo)scalar-diquark form factors: $F^+(Q^2)$ -- solid red curve; and  $F^-(Q^2)$ -- dashed blue curve.  There is practically no sensitivity to the DqAMM in these $J=0$ systems.
}
\end{figure}

Using the SCI, all photon+diquark vertices can be calculated following the pattern described in Ref.\,\cite{Roberts:2011wy}.  Herein, therefore, we will only present the results.  To begin, the elastic $\gamma \, 0^{\pm} \to 0^\pm$ vertices take the following form ($2K=p_f+p_i$):
\begin{equation}
\Lambda_\mu^\pm(p_f,p_i) = K_\mu F^{0^\pm}(Q^2)\,.
\end{equation}
The scalar functions can be computed; and on the domain $Q^2\in [0,10\,{\rm GeV}^2]$ they are accurately interpolated using the following $[1,2]$ Pad\'e approximant:
\begin{equation}
F^{0^\pm}(s=Q^2) =
\frac{a_0^\pm + a_1^\pm s  }
{1+b_1^\pm s + b_2^\pm s^2 }\,,
\label{Pade1}
\end{equation}
with the interpolation coefficients listed in Table~\ref{table-diquarkint}.  The results are drawn in Fig.\,\ref{fig:0pm0pmDFF}.

\begin{table}[b]
\caption{\label{table-diquarkint}
Interpolation coefficients to be used in Eq.\,\eqref{Pade1} for each respective elastic photon+diquark form factor.  All form factors are dimensionless, so each coefficient has the mass-dimension required to cancel that of the associated $Q^2\,({\rm GeV}^2)$ factor.}
\begin{tabular}{ccrrrr}
Process & $F_i(s)$ & $a_0$ & $a_1$ &  $b_1$ & $b_2$ \\
\hline
\tstrut
$\gamma 0^+ \to 0^+$ & $F_1$ & $1.000$ & $0.263$ & $1.402$ & $0.000$ \\[1ex]
$\gamma 0^- \to 0^-$ & $F_1$ & $1.000$ & $0.236$ & $1.611$ & $0.438$ \\[1ex]
$\gamma 1^+ \to 1^+$ & $F_1$ & $ 1.000$ & $ 2.211$ & $3.933$ & $3.052$ \\
                          & $F_2$ & $-2.869$ & $-0.044$ & $1.866$ & $0.007$ \\
                          & $F_3$ & $ 0.905$ & $ 0.362$ & $2.115$ & $3.744$ \\[1ex]
$\gamma 1^- \to 1^-$ & $F_1$ & $ 1.000$ & $ 0.099$ & $1.529$ & $0.417$ \\
                          & $F_2$ & $-2.623$ & $-2.071$ & $2.375$ & $1.790$ \\
                          & $F_3$ & $ 0.615$ & $-0.028$ & $1.458$ & $0.227$ \\\hline
\end{tabular}
\end{table}

\begin{figure}[t]
\vspace*{1ex}

\rightline{\hspace*{0.5em}{\large{\textsf{A}}}}
\vspace*{-3ex}
\includegraphics[width=0.42\textwidth]{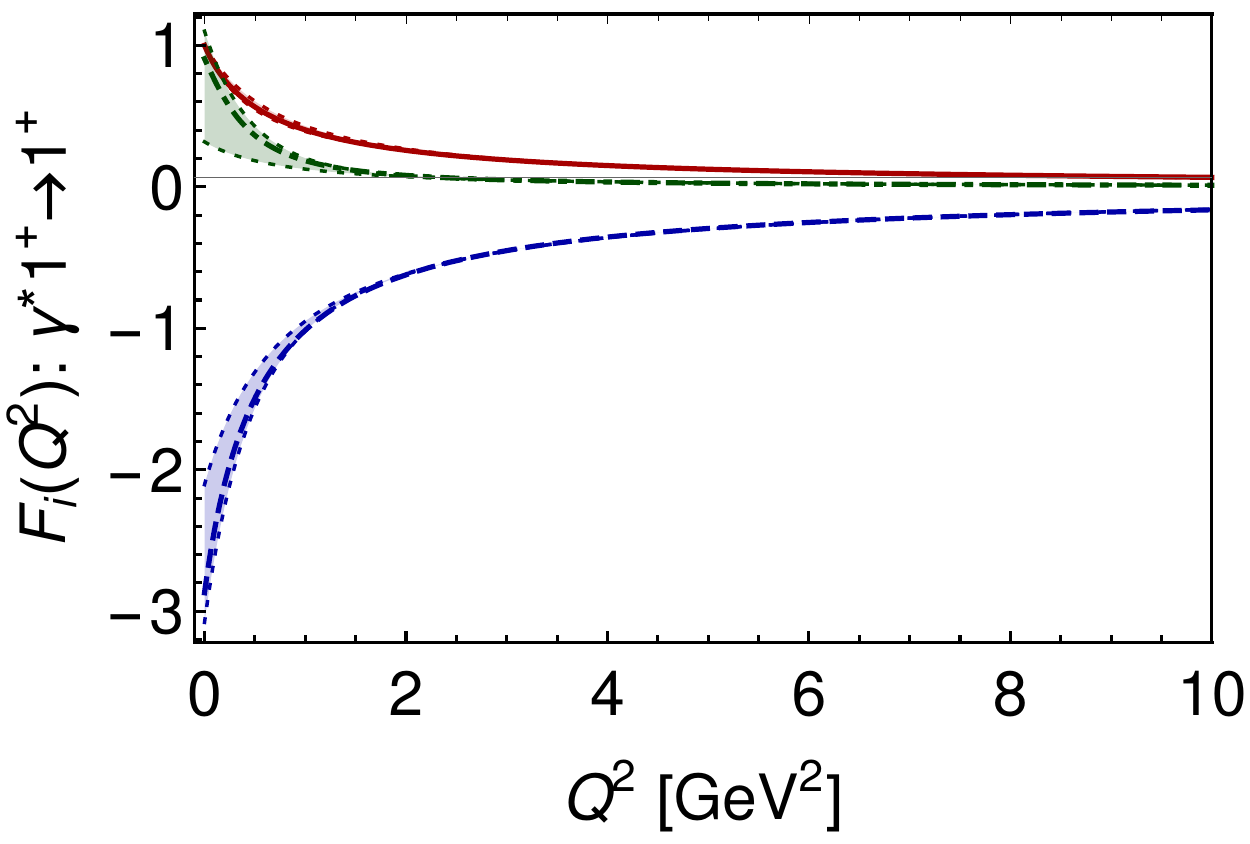}
\vspace*{1ex}

\rightline{\hspace*{0.5em}{\large{\textsf{B}}}}
\vspace*{-3ex}
\includegraphics[width=0.42\textwidth]{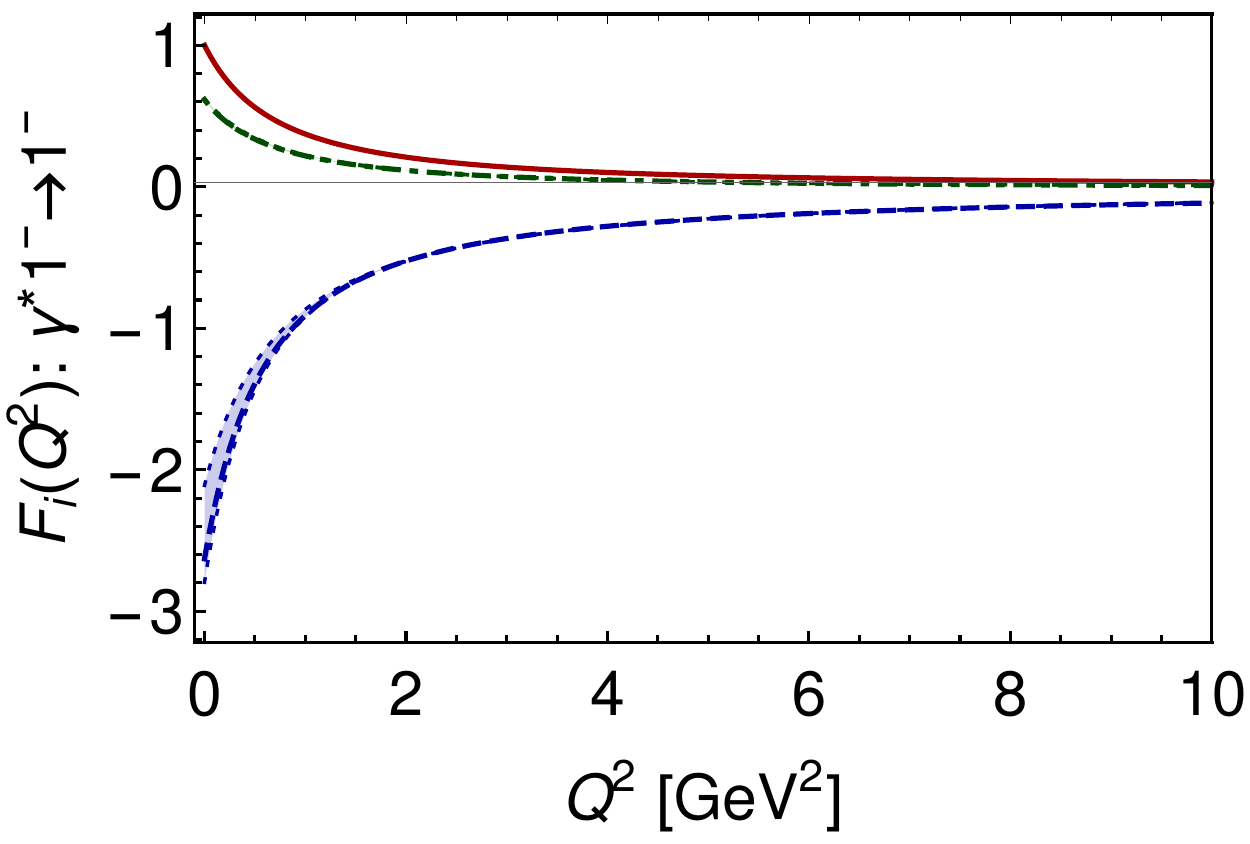}

\caption{\label{fig:1p1pDFF}
\emph{Upper panel}\,--\,{\sf A}.
Elastic photon+pseudovector-diquark form factors: $F_1(Q^2)$ -- solid red; $F_2(Q^2)$ -- dashed blue; and $F_3(Q^2)$ -- dot-dashed green.  In each case, the shaded areas show the response to variation of the DqAMM strength, $\zeta \in [0,0.5]$, around the highlighted $\zeta=1/3$ curves.
\emph{Lower panel}\,--\,{\sf B}.
Elastic photon+vector-diquark form factors with legend as in {\sf A}.
}
\end{figure}

Elastic electromagnetic form factors involving $1^\pm$ diquark correlations can be expressed as follows:
\begin{equation}
\Lambda_{\mu \alpha \beta}^\pm(p_i,p_f) = \sum_{j=1}^3 T_{\mu \alpha \beta}^{(j)}(K,Q)\, F_j^\pm(Q^2) \,,
\end{equation}
where
{\allowdisplaybreaks
\begin{align}
\label{eq:T111}
 T_{\mu \alpha \beta}^{(1)}(K,Q) & = 2 K_\mu\, {\mathpzc
 P}_{\alpha\sigma}(p_i) \, {\mathpzc P}_{\sigma\beta}(p_f)\,,  \\
 \label{eq:T211}
 T_{\mu \alpha \beta}^{(2)}(K,Q) & =  \left[Q_\alpha - p^i_{\alpha} \frac{Q^2}{2 m_{1^{\pm}}^2}\right] {\mathpzc P}_{\mu\beta}(p_f) \nonumber \\
 & \quad - \left[Q_\beta + p^f_\beta \frac{Q^2}{2 m_{1^{\pm}}^2}\right] {\mathpzc P}_{\mu\alpha}(p_i)\,, \\
 \label{eq:T311}
 T_{\mu\alpha \beta}^{(3)}(K,Q) & =  \frac{K_\mu}{m_{1^{\pm}}^2}\, \left[Q_\alpha - p^i_{\alpha} \frac{Q^2}{2 m_{1^{\pm}}^2}\right] \nonumber \\
 &\quad \times \left[Q_\beta + p^f_{\beta} \frac{Q^2}{2 m_{1^{\pm}}^2}\right] \,.
\end{align}
Our calculated results are accurately interpolated using a $[1,2]$ Pad\'e approximant of the form in Eq.\,\eqref{Pade1} with the coefficients in Table~\ref{table-diquarkint}.  Depicted in Fig.\,\ref{fig:1p1pDFF}, the form factors are similar to those of a vector meson \cite{Roberts:2011wy} and, as in that case, the magnetic form factor $G_M = -F_2$.  The DqAMM has an observable impact on the elastic electromagnetic form factors of these $J=1$ systems.}

\subsection{Photon-induced diquark transition vertices -- S3}
\label{ES3}
Diagram S3 in Fig.\,\ref{fig:current} represents five electromagnetically induced diquark transition vertices:
scalar$\,\leftrightarrow\,$pseu\-dovector; pseudoscalar$\,\leftrightarrow\,$vector;
scalar$\,\leftrightarrow\,$vector; pseu\-do\-scalar$\,\leftrightarrow\,$pseudovector;
and pseudovector$\,\leftrightarrow\,$vec\-tor.  \linebreak

\begin{figure}[t]
\begin{tabular}{c}
\includegraphics[width=0.42\textwidth]{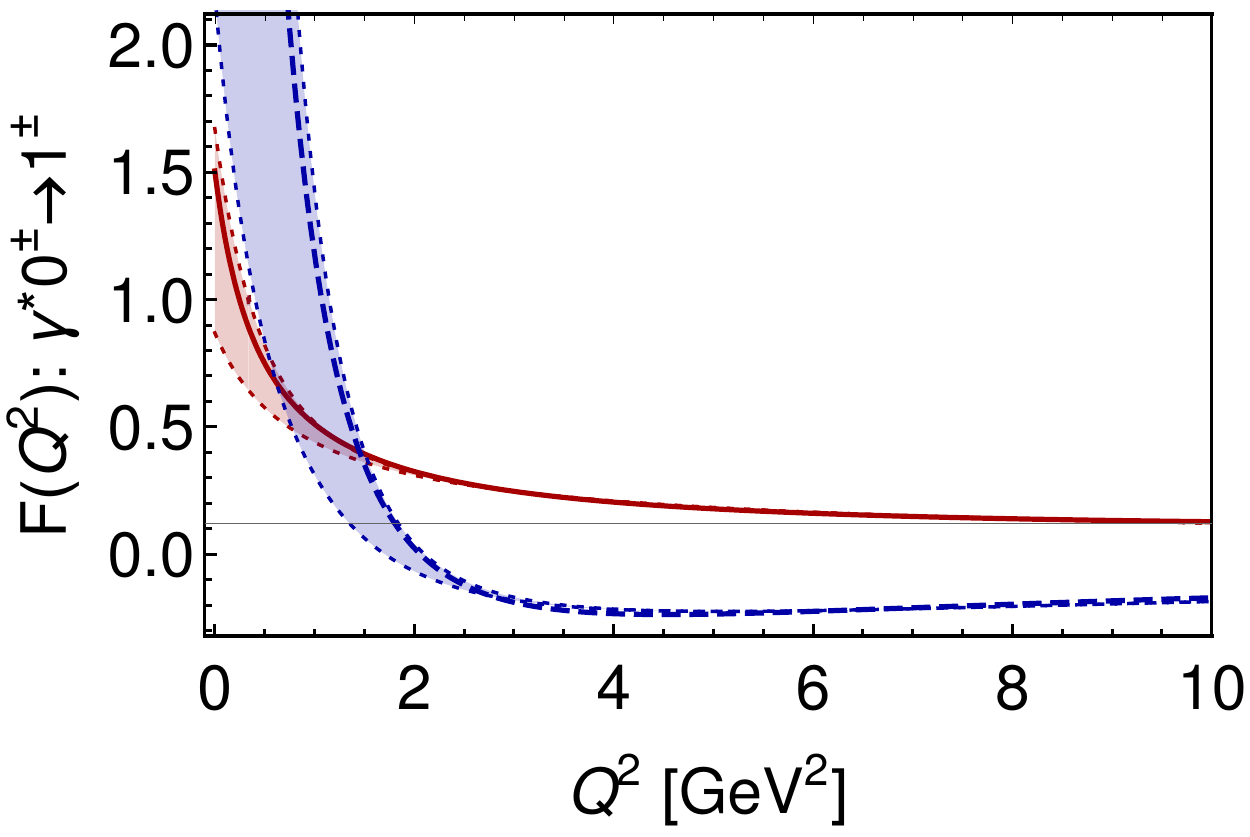} \\
\end{tabular}
\caption{Form factors for photon induced transition between: 
scalar-diquark and pseudovector-diquark, $\gamma^{(\ast)} 0^+ \to 1^+$ -- solid red curve; and pseudoscalar-diquark and vector-diquark, $\gamma^{(\ast)} 0^- \to 1^-$ -- dashed blue curve.
In each case, the shaded areas show the response to variation of the DqAMM strength, $\zeta \in [0,0.5]$, around the highlighted $\zeta=1/3$ curves.
\label{fig:1pm0pmDFF}}
\centering
\end{figure}

The first two involve like-parity diquarks in the initial and final states and have the following simple structure:
\begin{equation}
\Lambda_{\mu \rho}^{1^\pm 0^\pm}(p_f,p_i) = \frac{1}{m_1^\pm}\epsilon_{\mu\rho\alpha\beta} Q_\alpha p_{f\beta} F^\pm(Q^2)\,.
\label{eq:VSep}
\end{equation}
Our calculated results are accurately interpolated by a $[1,2]$ Pad\'e approximant in the form of Eq.\,\eqref{Pade1} with the coefficients given in Table~\ref{table-diquarkint2}.  They are drawn in Fig.\,\ref{fig:1pm0pmDFF}.  Plainly, the DqAMM has a noticeable impact on both vertices and the $\gamma 0^- \to 1^-$ transition form factor is typically larger in magnitude than that describing $\gamma 0^+ \to 1^+$.  The latter feature has little impact, however, because the nucleon contains practically no negative-parity diquarks.

\begin{table}[b]
\caption{\label{table-diquarkint2}
Interpolation coefficients to be used in Eq.\,\eqref{Pade1} for each photon-induced diquark transition form factor.
There are two exceptions: $[2,3]$ Pad\'e approximants are required to accurately represent the $\gamma^{(\ast)} 1^+ \to 0^-$, $\gamma^{(\ast)} 0^+ \to 1^-$ transitions.  In each of these cases there are two additional coefficients for both of the functions involved, Eq.\,\eqref{eq:VSdp}.
$\gamma^{(\ast)} 1^+ \to 0^-$: $a_2 = 0.267$, $b_3 = 0.029$ ($F_1$); and $a_2 = -0.019$, $b_3 = 0.348$ ($F_2$).
$\gamma^{(\ast)} 1^- \to 0^+$: $a_2 = 0.227$, $b_3 = 1.852$ ($F_1$); and $a_2 = 0.051$, $b_3 = 2.449$ ($F_2$).
All form factors are dimensionless, so each coefficient has the mass-dimension required to cancel that of the associated $Q^2\,({\rm GeV}^2)$ factor.}
\begin{tabular}{ccrrrr}
Transition & $F_i(s)$ & $a_0$ & $a_1$ &  $b_1$ & $b_2$ \\
\hline
\tstrut
$\gamma 0^+ \to 1^+$ & $F_1$ & $1.505$ & $0.069$ & $2.134$ & $-0.051$ \\[1ex]
$\gamma 0^- \to 1^-$ & $F_1$ & $12.852$ & $-6.237$ & $1.926$ & $2.691$\\[1ex]
$\gamma 1^+ \to 0^-$ & $F_1$ & $0.138$ & $0.097$ & $0.368$ & $1.769$ \\
                          & $F_2$ & $0.066$ & $0.104$ & $1.604$ & $1.740$ \\[1ex]
$\gamma 1^- \to 0^+$ & $F_1$ & $-0.465$ & $4.657$ & $4.689$ & $5.365$ \\
                          & $F_2$ & $-0.347$ & $2.663$ & $4.739$ & $5.006$ \\[1ex]
$\gamma 1^- \to 1^+$ & $F_1$ & $ 1.729$ & $ 0.284$  & $1.253$ & $-0.028$ \\
                          & $F_2$ & $-0.226$ & $ 0.813$  & $1.074$ & $-0.031$ \\
                          & $F_3$ & $-3.397$ & $-17.375$ & $6.329$ & $ 6.211$ \\\hline
\end{tabular}
\end{table}

The next two transitions involve opposite parity diquarks: $\gamma 1^\pm \to 0^\mp $. They are characterised by two form factors \cite{Dudek:2009kk}:
\begin{equation}
\label{eq:VSdp}
\Lambda_{\mu \rho}^{\mp \pm}(p_i,p_f) = \sum_{j=1}^2 T_{\mu\rho}^{(j)}(p_i,p_f) F_j^{\mp\pm}(Q^2) \,,
\end{equation}
where
\begin{subequations}
\begin{align}
T_{\mu\rho}^{(1)}(p_i,p_f) &= \frac{m_{1^\pm}}{\mathcal{D}} \big[\Omega\;\mathcal{P}^{T}_{\mu\rho}(p_f) \nonumber \\
&
\hspace*{-0.60cm} -\mathcal{P}^{T}_{\mu\nu}(p_f)p_{i\nu}(p_{f\rho}(p_f\cdot p_i)-m_{1^\pm}^2 p_{i\rho}) \big] \,, \\
T_{\mu\rho}^{(2)}(p_i,p_f) &= \frac{m_{1^\pm}}{\mathcal{D}} \mathcal{P}^{T}_{\mu\nu}(p_f)p_{i\nu}\big[(p_f\cdot p_i)(p_f+p_i)_{\rho}  \nonumber \\
&
\hspace*{-0.60cm} - m_{0^\pm}^{2}p_{f\rho}-m_{1^\pm}^{2}p_{i\rho} \big] \,,
\end{align}
\end{subequations}
with $\Omega = (p_i \cdot p_f)^2- p_i^2 p_f^2$ and $\mathcal{D} = \Omega, m_{1^-}^4$ for the $\gamma  1^+ \to 0^-$  and $\gamma 1^-\to 1^+ $ cases, respectively.

\begin{figure}[t]
\vspace*{1ex}

\rightline{\hspace*{0.5em}{\large{\textsf{A}}}}
\vspace*{-3ex}
\includegraphics[width=0.42\textwidth]{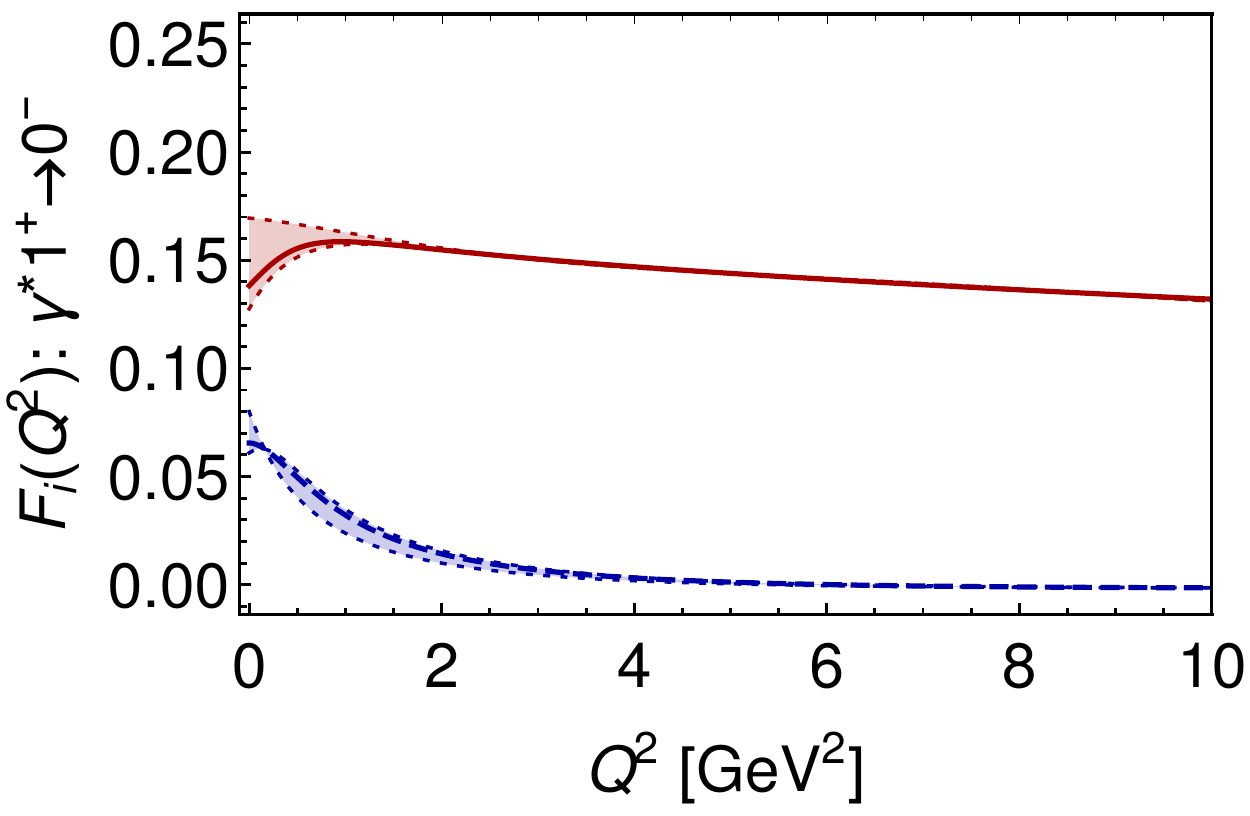}
\vspace*{1ex}

\rightline{\hspace*{0.5em}{\large{\textsf{B}}}}
\vspace*{-3ex}
\includegraphics[width=0.42\textwidth]{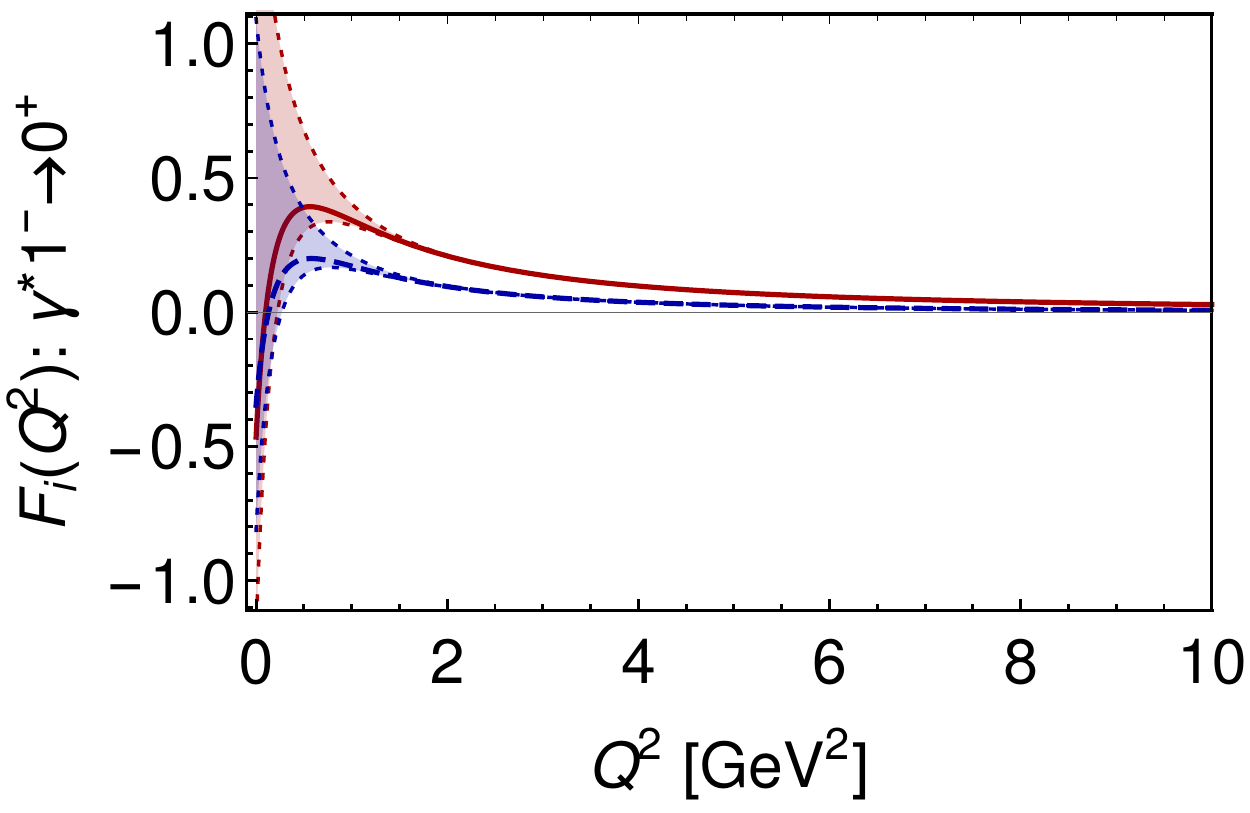}

\caption{
\emph{Upper panel}\,--\,{\sf A}.
$\gamma^{(\ast)} 1^+ \to 0^-$ transition form factors, Eq.\,\eqref{eq:VSdp}: $F_1(Q^2)$ -- solid red; and $F_2(Q^2)$ -- dashed blue.
\emph{Lower panel}\,--\,{\sf B}.
$\gamma^{(\ast)} 1^- \to 0^+$ transition form factors with  legend as in {\sf A}.
In each case, the shaded areas show the response to variation of the DqAMM strength, $\zeta \in [0,0.5]$, around the highlighted $\zeta=1/3$ curves.
\label{fig:1p0mDFF}}
\centering
\end{figure}

Accurate interpolations of the computed results for these form factors are provided by $[2,3]$ Pad\'e approximants with the coefficients described in Table~\ref{table-diquarkint2}.  They are illustrated in Fig.\,\ref{fig:1p0mDFF}.  Evidently, $F_1$ is dominant in both cases and the $0^+\leftrightarrow 1^-$ transition form factors exhibit greater sensitivity to the DqAMM.

The final transition is $\gamma 1^- \to 1^+$, a complete description of which requires three form factors:
{\allowdisplaybreaks
\begin{equation}
\Lambda_{\mu \alpha\beta}^{1^+1^-}(K,Q) =  \sum_{j=1}^3 T_{\mu \alpha\beta}^{(j)}(K,Q) \, F_j^{1^+1^-}(Q^2) \,, \\
\label{Lambd1p1m}
\end{equation}
where
\begin{subequations}
\begin{align}
T_{\mu \alpha\beta}^{(1)}(K,Q) &= m_{1^-} \, \frac{\epsilon_{\mu\rho\sigma\gamma}(p_{i}-p_{f})_{\gamma}} {4\sqrt{2}\Omega} \nonumber \\
&
\times (p_{i}+p_{f})_{\sigma} \, \big[2m_{1^-} \,  \mathcal{P}_{\lambda\alpha}^{\bot}(p_{i}) \, p^{f}_{\lambda} \, \mathcal{P}_{\rho \beta}^{\bot}(p_{f}) \nonumber \\
&
+2m_{1^+} \, \mathcal{P}_{\lambda\beta}^{\bot}(p_{f}) \, p^{i}_{\lambda} \,  \mathcal{P}_{\rho\alpha}^{\bot}(p_{i}) \big] \,, \\
T_{\mu \alpha\beta}^{(2)}(K,Q) &= m_{1^-} \, \frac{\epsilon_{\mu\rho\sigma\gamma}(p_{i}-p_{f})_{\gamma}} {4\sqrt{2}\Omega} \nonumber \\
&
\times (p_{i}+p_{f})_{\sigma} \big[2m_{1^-} \,  \mathcal{P}_{\lambda\alpha}^{\bot}(p_{i}) \, p^{f}_{\lambda} \, \mathcal{P}_{\rho\beta}^{\bot}(p_{f}) \nonumber \\
&
-2m_{1^+} \,  \mathcal{P}_{\lambda\beta}^{\bot}(p_{f}) \, p^{i}_{\lambda} \,  \mathcal{P}_{\rho\alpha}^{\bot}(p_{i}) \big] \,, \\
T_{\mu \alpha\beta}^{(3)}(K,Q) &= \frac{\epsilon_{\mu\rho\sigma\gamma}(p_{i}-p_{f})_{\gamma}} {4\sqrt{2}\Omega}  \nonumber \\
&
\hspace{-0.8cm} \times \big(-4\Omega \, \mathcal{P}_{\rho\alpha}^{\bot}(p_{i}) \, \mathcal{P}_{\sigma\beta}^{\bot}(p_{f}) \nonumber \\
&
\hspace{-0.8cm} + (p_{i}+p_{f})_{\sigma} [(p_{i}^{2}-p_{f}^{2}+Q^{2}) \, p_{f \lambda} \, \mathcal{P}_{\lambda\alpha}^{\bot}(p_{i}) \, \mathcal{P}_{\rho\beta}^{\bot}(p_{f}) \nonumber \\
&
\hspace{-0.8cm} + (p_{i}^{2}- p_{f}^{2}-Q^{2}) \, p_{i \lambda} \, \mathcal{P}_{\lambda\beta}^{\bot}(p_{f}) \,  \mathcal{P}_{\rho\alpha}^{\bot}(p_{i})]\big) \;.
\end{align}
\end{subequations}
This case requires that one evaluate the two possible orderings of incoming/outgoing diquarks in order to guarantee that the vertex $\Lambda_{\mu \alpha\beta}^{1^+1^-}(K,Q)$ is symmetric under the simultaneous interchanges $p_i \leftrightarrow p_f$, $\alpha \leftrightarrow \beta$.
}

\begin{figure}[!t]
\includegraphics[width=0.45\textwidth]{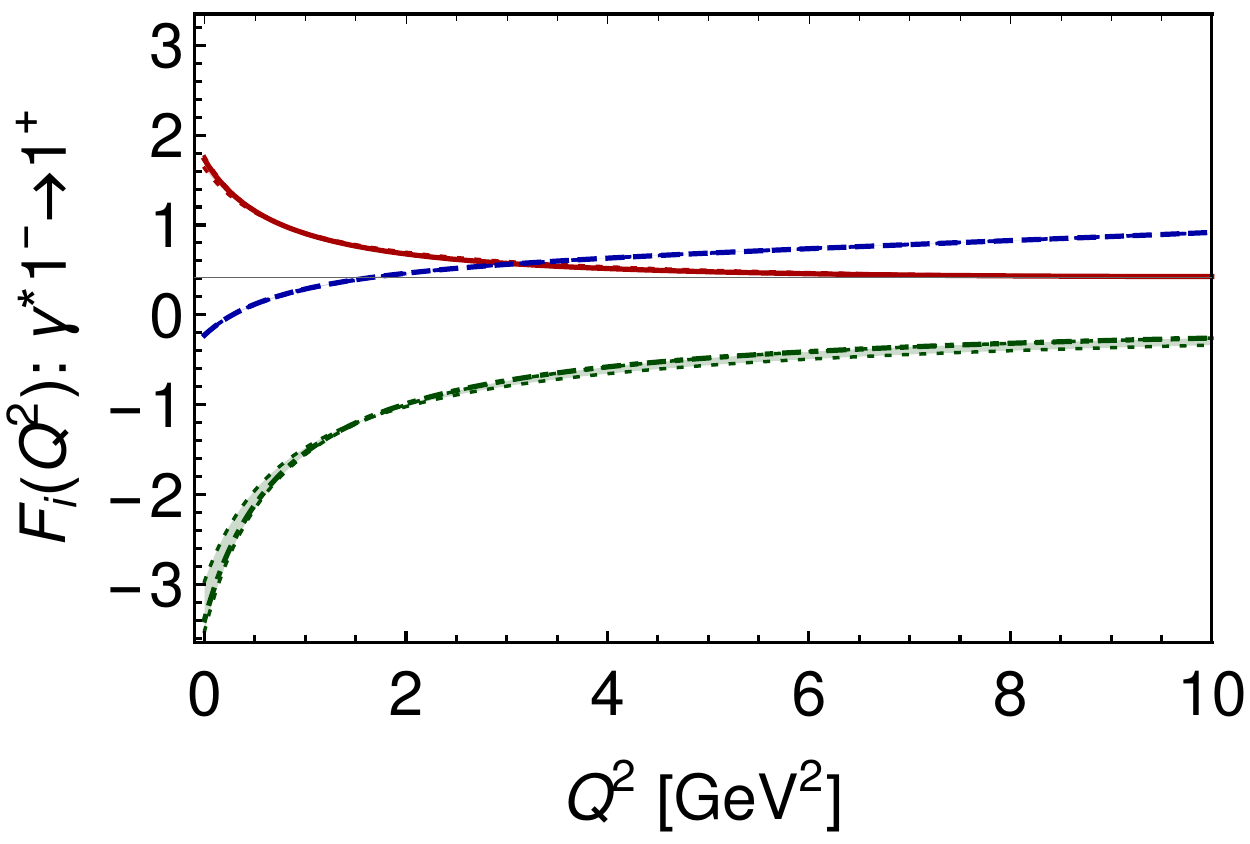}
\caption{Photon induced $1^+ \leftrightarrow 1^-$ transition form factors, Eq.\,\eqref{Lambd1p1m}:
$F_1(Q^2)$ -- solid red; $F_2(Q^2)$ -- dashed blue; and $F_3(Q^2)$ -- dot-dashed green The shaded areas show the variation of the strength of the AMM, $\zeta \in [0,0.5]$; the highlighted curves correspond to $\zeta=1/3$.
The shaded band shows the response to variation of the DqAMM strength, $\zeta \in [0,0.5]$, around the highlighted $\zeta=1/3$ curves.
\label{fig:1m1pDFF}}
\end{figure}

The calculated form factors are drawn in Fig.\,\ref{fig:1m1pDFF}.  Accurate interpolations of the results are provided by $[1,2]$ Pad\'e approximants with the coefficients listed in Table~\ref{table-diquarkint2}.  The DqAMM has only a marginal impact on these transition form factors.

\setcounter{equation}{0}

\section{Nucleon elastic and transition form factors}
\label{apptranscurr}
In our SCI quark+diquark picture of baryons, each elastic and transition form factor can be divided into two separate contributions: photon strikes quark; and photon strikes diquark.  Thus, one may rewrite Eq.\,\eqref{JNNastExplicit} as follows:
\begin{align}
\nonumber
& \Gamma_\mu^{BA}(P_f,P_i) \\
& = \sum_{I=S1,S2,S3}\int_l \Lambda_+^B(P_f) \Lambda_\mu^I(l;P_f,P_i) \Lambda_+^A(P_i)\,, \nonumber \\
& =: \int_l \Lambda_+^B(P_f)
\left[
 \sum_r \mathcal{Q}_\mu^{(j)} + \sum_{s,t} \mathcal{D}_\mu^{(s,t)}
\right]
\Lambda_+^A(P_i)
\label{JNNastExplicitB}
\end{align}
where $BA= ++$, $--$, $-+$, as before, $\int_l$ is our SCI regularisation of the four-dimensional integral;
and
$\mathcal{Q}_\mu^{(r)}$ is a diagram in which the photon strikes a quark with a diquark spectator, labelled by $r=0^+,1^+,0^-,1^-$,
whereas
$\mathcal{D}_\mu^{(s,t)}$ indicates a diagram with a quark spectator to a diquark interaction $s\leftrightarrow t$, $s,t=0^+,1^+,0^-,1^-$.

\subsection{Photon strikes quark}
\label{Appgdq}
This contribution has the general form
\begin{eqnarray}
\label{eq:PHQ}
\mathcal{Q}^{(r)}_\mu &=& q_r \int_l \bar{\psi}_{(m)}^{f(r)} S(l_f^+)\Gamma_\mu^\gamma (Q) S(l_i^+) \psi_{(m)}^{i(r)} \Delta^{(r)}(-l),
\end{eqnarray}
where $l_{f,i}^\pm = \pm l + P_{f,i}$,
$P_{f,i}^2=-M_{f,i}^2$, with $M_{f,i}$ being the masses of the baryons involved,
and $Q^2=(P_f-P_i)^2$.
Here, referring to Eq.\,\eqref{nucleonamplitude}, $\psi_{(m)}^{i(r)}$ denotes that part of the Faddeev amplitude for the indicated baryon that is associated with component-$m$ of the diquark type $r$ bystander; and $q_r$ is the charge of the struck quark in units of the positron charge.  Depending on $r$, the diquark propagator may have Lorentz indices that are contracted with those of the Faddeev amplitude, also suppressed:
\begin{eqnarray}
\label{eq:diqprop1}
\Delta^{0^\pm}(K) &=& \frac{1}{K^2+m_{0^\pm}^2}\;,\\
\label{eq:diqprop2}
\Delta^{1^\pm}_{\mu\nu}(K) &=& \frac{1}{K^2+m_{1^\pm}^2}\left(\delta_{\mu\nu}+\frac{\delta_\mu \delta_\nu}{m_{1^\pm}^2} \right)\,.
\end{eqnarray}

It is worth providing some details here on the $\Gamma_\mu^{-+}=\gamma^{(\ast)} p \to N^\ast(1535)\,\tfrac{1}{2}^-$ transition.
Suppose $r=0^+$, then $q_r=2/3$, $m=1$, and
\begin{equation}
\psi_{(1)}^{i (0^+)} = {\mathbf I}_{\rm D} {\mathpzc s}^{+}\,,\quad
\bar\psi_{(1)}^{f (0^+)} = i {\mathpzc p}^- {\mathbf I}_{\rm D} \,.
\end{equation}
The $r=0^-$ case is obvious by analogy.

Consider next the case $r=1^+$.  Then $m=1,2$,
\begin{subequations}
\begin{align}
\psi_{\beta (1)}^{i(1^+){\mathpzc j}} & = a_1^{+{\mathpzc j}} \gamma_5\gamma_\beta \,,
\quad \psi_{\beta (2)}^{i(1^+){\mathpzc j}} = a_2^{+{\mathpzc j}} \gamma_5 \hat P_\beta\,, \\
\bar\psi_{\alpha(1)}^{f(1^+){\mathpzc j}} & = - a_1^{-{\mathpzc j}}\gamma_\alpha \,, \quad
\bar\psi_{\alpha(2)}^{f(1^+){\mathpzc j}} = -a_2^{-{\mathpzc j}}{\mathbf I}_{\rm D} \hat P_\alpha  \,.
\end{align}
\end{subequations}
In these expressions:
${\mathpzc j}=1 \Rightarrow \{uu\}_{1^+}$, which means $q_{1^+}^1=-1/3$; and
${\mathpzc j}=2 \Rightarrow \{ud\}_{1^+}$, $q_{1^+}^2=2/3$.
We work in the isospin-symmetry limit; so, as noted following Eq.\,\eqref{spav},
${\mathpzc a}^{{\mathpzc j}=2}=-{\mathpzc a}^{{\mathpzc j}=1}/\surd 2$.  Hence, the terms in this $1^+$-spectator contribution combine as follows:
\begin{equation}
q_{1^+}^1 \, a_1^{+{\mathpzc j}=1} \, a_1^{-{\mathpzc j}=1}
+ q_{1^+}^2 \, a_1^{+{\mathpzc j}=2} \, a_1^{-{\mathpzc j=2}} = 0\,,
\end{equation}
etc.  Namely, they cancel.

Following this pattern, the contribution to Eq.\,\eqref{JNNastExplicitB} connected with the vector diquark bystander is readily constructed.  Since this is an isoscalar diquark, there is no cancellation in this case.

Consider now Eq.\,\eqref{eq:amplitudes}.
Observe that the initial state nucleon has practically no pseudoscalar or vector diquark content; and we have just seen that the contributions from pseudovector-diquark bystanders cancel amongst themselves.
Consequently, regarding the $\gamma^{(\ast)} p \to N^\ast(1535)\,\tfrac{1}{2}^-$ transition, only the $0^+$ diquark spectator diagram can make a material contribution.

The calculation of any given contribution is completed by using a Feynman parametrisation to combine denominators, followed by evaluation of the four-dimension\-al integral following usual SCI procedures.  An explicit example may be found in Ref.\,\cite{Wilson:2011aa}.

\subsection{Photon strikes diquark}
For this class of processes, the general expression is:
\begin{align}
\label{eq:PHD}
\mathcal{D}^{(s,t)}_\mu &= q_{ts} \int_l
\bar{\psi}^{f(t)}_{(m)} S(l) \psi^{i(s)}_{(m)}
\Delta^{t}(l_f^-) \Lambda_{\mu}(l_f^-,l_i^-) \Delta^{s}(l_i^-) \;,
\end{align}
where $\Lambda_\mu$ corresponds to the appropriate photon-diquark vertex in \ref{App:current} and, as above, the diquark propagator and Faddeev amplitude component may have contracted Lorentz indices.

Focusing again on $\gamma^{(\ast)} p \to N^\ast(1535)\,\tfrac{1}{2}^-$, using the $\gamma 0^+ \to 1^+$ transition as an example and referring to Eq.\,\eqref{nucleonamplitude}, one has $q_{ts}=q_{\{ud\}}=1/3$,
\begin{subequations}
\begin{align}
\psi^{i(0^+)}_{(1)} & ={\mathbf I}_{\rm D} {\mathpzc s}^+\,,
\;
\bar{\psi}_{(1,2)\beta}^{f(1^+)} =
 -( \mathpzc{a}_1^- \gamma_\beta + \mathpzc{a}_2^- \hat{P}_\beta ) \\
\Delta^{s} & = \Delta^{0^+}\,,
\; \Delta^{t} = \Delta^{1^+}_{\beta\alpha}\,,
\; \Lambda_\mu = \Lambda_{\mu \alpha}^{1^+0^+}\,,
\end{align}
\end{subequations}
where $\Lambda_{\lambda\alpha}^{1^+0^+}$ is given in Eq.\,\eqref{eq:VSep}.  All other cases are equally straightforward.

Again, the evaluation of any given contribution is completed by following the procedures established in Ref.\,\cite{Wilson:2011aa}.


\end{document}